\documentclass[prd,10pt,twocolumn,showpacs,nofootinbib,superscriptaddress,amsmath,amssymb]{revtex4-1}

\usepackage{graphicx}% Include figure files
\usepackage{dcolumn}% Align table columns on decimal point
\usepackage{bm}% bold math

\usepackage{setspace}
\usepackage[T1]{fontenc} % if needed
\usepackage[varg]{txfonts}
\usepackage[]{natbib}
\usepackage[english]{babel}
\usepackage{amsmath}
\usepackage{ragged2e}
\usepackage{amssymb}
\usepackage{hyperref}
\usepackage{ccaption}
\usepackage{longtable}
\usepackage{graphics,psfrag,rotating}
\usepackage{graphicx}% Include figure files
\usepackage{bm}% bold math
\usepackage{color}
\usepackage{float}
\usepackage[dvipsnames,table]{xcolor}
\hypersetup{urlcolor=RedViolet,
	    citecolor=ForestGreen ,
	    linkcolor=RoyalBlue, 
	    colorlinks=true}
\definecolor{lightgray}{rgb}{0.9,0.9,0.9}	    
\definecolor{green}{rgb}{0,0.5,0}
\definecolor{red}{rgb}{1,0,0}
\definecolor{blue}{rgb}{0,0,0.5}

\newcommand{\lsim}   {\mathrel{\mathop{\kern 0pt \rlap
  {\raise.2ex\hbox{$<$}}}
  \lower.9ex\hbox{\kern-.190em $\sim$}}}
\newcommand{\gsim}   {\mathrel{\mathop{\kern 0pt \rlap
  {\raise.2ex\hbox{$>$}}}
  \lower.9ex\hbox{\kern-.190em $\sim$}}}

\newcommand{\blue}{\color{blue}}
\newcommand{\aap}{{Astron.~Astrophys.}}

\newcommand{\physrep}{{Phys.~Rep.}}

\definecolor{myred}{RGB}{220, 20, 60}
\definecolor{myblue}{RGB}{60, 20, 220}
\definecolor{mygreen}{RGB}{60, 220, 20}

\begin{document}

\preprint{APS/123-QED}

\title{Theoretical predictions for dark matter detection in dwarf irregular galaxies
with gamma rays}% Force line breaks with \\

\author{V. Gammaldi}\email{vgammald@sissa.it}
 \affiliation{SISSA, International School for Advanced Studies, \\
  Via Bonomea 265, 34136, Trieste, Italy}
 \affiliation{INFN, Istituto Nazionale di Fisica Nucleare - Sezione di Trieste, \\
 Via Valerio 2, 34127, Trieste, Italy}

\author{E. Karukes}
\email{ekarukes@ift.unesp.br}
\affiliation{ICTP-SAIFR \& IFT-UNESP, \\
R. Dr. Bento Teobaldo Ferraz 271, S\~ao Paulo, Brazil}

\author{P. Salucci}
\email{salucci@sissa.it}
 \affiliation{SISSA, International School for Advanced Studies, \\
  Via Bonomea 265, 34136, Trieste, Italy}
 \affiliation{INFN, Istituto Nazionale di Fisica Nucleare - Sezione di Trieste, \\
 Via Valerio 2, 34127, Trieste, Italy}

\date{\today}% It is always \today, today,
\smallskip
\begin{abstract}
We investigate rotationally supported dwarf irregular (dIrr) galaxies as a new category of targets for indirect Dark Matter (DM) searches with gamma-ray telescopes. In the framework of point-like analysis, pressure-supported dwarf spheroidal (dSph) galaxies are usually considered as one of the best category of targets for indirect DM searches, due to their close distance and negligible astrophysical background. Nonetheless, as a result of their uncertain kinematics, the DM content and astrophysical J-factors of dSphs are usually affected by significant errors.
In this paper, we study a sample of 36 dIrrs as prospective targets of interest. In the framework of the universal rotation curve, the kinematics of dIrr galaxies provides a good estimation of their DM halo density distribution and, consequently, of their astrophysical J-factors. We calculate the J-factors for these 36 dIrr galaxies, whose kinematics have been studied in a previous work. We find a range of values 
comparable with the J-factors of dSph galaxies.
However, differently from dSphs an \textit{extra} astrophysical gamma-ray background component is expected in dIrr galaxies, that is due to their star-formation activity. In this paper, we show via a theoretical approach that for galaxies in our sample the \textit{extra} astrophysical background component is negligible. Therefore, we conclude that dIrr galaxies can be potentially considered as additional point-like targets for DM searches with gamma-rays. As a first application of this study, we show the sensitivity limits of the Fermi-LAT telescope to these objects and we calculate constraints on the DM particle mass and annihilation cross-section.  We conclude that the results of the individual study of several dIrr galaxies are not yet competitive with respect to the analysis of one of the most promising dSph galaxies, i.e. SEGUE1. However, taking into account SEGUE1's symmetry-related uncertainties in the J-factor calculation might alter this conclusion.  Additionally, we calculate constraints for the combined analysis of the 7 most promising dIrr galaxies of our sample.
\begin{description}
\item[PACS numbers]
\end{description}
\end{abstract}

\pacs{Valid PACS appear here}% PACS, the Physics and Astronomy
\maketitle

%\tableofcontents

\section{Introduction}

Astrophysical and cosmological evidences suggest that non-baryonic cold dark matter (DM) constitutes $84\%$ of the matter density of the Universe \citep{Ade:2015xua}. Many well-motivated 
 DM candidates are predicted to annihilate or decay into Standard Model (SM) particles and to produce secondary cosmic rays, such as gamma rays \citep{Bertone:2005xv}.  
Galaxy clusters, dwarf spheroidal (dSph) galaxies as well as the Galactic Center are principal astrophysical targets for indirect DM searches. 
Among other targets, the Milky Way dSph galaxies are considered to be especially promising objects due to their relatively close position and their appearance as point-like or marginally extended sources in gamma-ray telescopes. Moreover, the contamination from intrinsic astrophysical sources is negligible in these objects. In fact, they host an old stellar population of low-luminosity  and do not possess gas. However, the high uncertainties in the kinematics of these pressure-supported galaxies and the total absence of gas in the periphery do not allow us to well constrain their DM profiles  \citep[see, e.g.,][]{dwarfkin}. The effective determination of the latter is very crucial for the estimation of the astrophysical factor (or J-factor), that depends on the DM density distribution profile. Besides,  due to the uncertainty on the geometry of dSphs, the astrophysical factor 
is affected by the errors up to six orders of magnitude \citep{triaxSph}.  
In spite of that, the study of dSph galaxies sets the most stringent constraints on the particle DM mass and the annihilation cross-section so far. In particular, the stacked analysis of several dSphs allows to exclude the thermal DM particle of the mass below 100 GeV \citep{dSphFERMI1}. \\

In this paper, driven by the intent of reducing the uncertainty on the astrophysical J-factor for DM indirect searches, we investigate a sample of 36 dwarf irregular (dIrr) galaxies of the Local Volume catalog \citep{karachentsev}, that is within a sphere of $\sim 11$ Mpc centred on our Galaxy and within redshift $z\lsim 10^{-3}$. Their distances obtained by means of primary distance indicators, are comparable with that of many dSph galaxies. Unlike pressure supported dSph galaxies, dIrrs are rotationally supported star-forming dwarf galaxies, that are considered to be DM dominated objects at all radii \citep{oh,oh1,gentile06}. Their kinematics is rather simple and points to the existence of cored DM density profiles with halos much larger than the enclosed star-forming regions \citep{urc1}. These relevant properties and the increasing number of dIrr galaxies that has been recently detected and studied in their kinematics, make them interesting targets for DM searches.\\ 

Nonetheless and unlike dSphs, an astrophysical gamma-ray contamination is expected due to the star-forming activity of these objects \citep{Martin}. Assuming that the main contribution is enclosed in the optical regions of these galaxies, in this paper we study the possibility of considering the big halos of dIrr galaxies as targets of interest, in the framework of point-like analysis for DM indirect searches with gamma-ray telescopes.\\

The paper is organized as follows. In Section \ref{Irr} we briefly review the DM density distribution profiles of our sample of dIrr galaxies. Details on gamma-ray detectors are given in Section \ref{telescopessection}. In Section \ref{gammarayDMsection}, we briefly review the fundamentals of DM searches with gamma rays. We discuss and calculate the astrophysical J-factors of the dIrr galaxies in our sample and the associated uncertainties in Section \ref{Jfactor}, including details on the baryon and DM distribution in dIrrs. In Section \ref{gammarayBg} we theoretically estimate  two contributions to the gamma-ray emission in dIrr, both from astrophysics and DM.
As a result of the sensitivity study, we present the constraints on the DM particle mass and annihilation cross-section by means of both the individual and combined analysis of galaxies in the sample in Section \ref{analysissecVII}. In Sections \ref{discussion} and \ref{conclusions} we discuss the results and summarise the main conclusions of this work. Further information about the rotation curves of our galaxy sample, details on the astrophysical J-factor and the sensitivity study for each individual galaxy are given in Appendices \ref{AppA} , \ref{AppB} and \ref{AppC}, respectively.

\section{Dwarf Irregular Galaxies and Burkert profile}
\label{Irr}

The sample of 36 dIrr galaxies was presented and analysed in \cite{Karukes}, where the concept of rotation curve universality is applied. 
This concept is based on the systematic study of the rotation curves of more than 1100 spiral galaxies  \citep{urc,urc1}.
The Universal Rotation Curve (URC) allows to describe the structural parameters of luminous and dark components in galaxies without individual mass modelling, and at the same time reduces the uncertainties on the estimation of the DM density distribution profile. As shown in \cite{Karukes} and in Appendix \ref{AppA} of this paper, the URC model allows to well fit the rotation curve for this class of objects, for which the few kinematical data do not allow an individual mass modelling. The DM distribution in these galaxies is well described by the Burkert profile:
\begin{equation}
\rho_{\mathrm{Burkert}}(r)=\frac{\rho_0r_0^3}{(r+r_0)(r^2+r_0^2)},
\label{Burkert}
\end{equation}
where $\rho_0$ is the central density and $r_0$ is the core radius. Let us notice, that the cuspy NFW profile \citep{Navarro1995} does not fit well the kinematics of dIrr galaxies \citep{deBlok2009,Karukes}. 
Instead, the URC model and the Burkert profile are able to well recover the observed kinematics of most of the galaxies in the sample. The uncertainties on the estimated inclinations or disk length scales of the objects could justify the three cases (NGC6822, AndIV and UGC8508) in which the URC model does not reproduce the outer kinematics. For this reason, we do not apply any selection to the sample, but we keep in mind that their observed kinematics should be farther checked. In Fig. \ref{v12} of Appendix~\ref{AppA}, we show the observed and modelled rotation curves of the full sample of 36 dIrrs galaxies. 
The DM halo parameters for the galaxies in the sample are listed in Table \ref{Tab:Irr}\footnote{The DM parameters  listed in Table \ref{Tab:Irr} are slightly different from the ones in \citet{Karukes} due to the fact that for some of the galaxies the values of the inclination were updated. Additionally, distances are also slightly different from the ones cited in \citet{karachentsev} due to the recent updates in the catalog http://www.sao.ru/lv/lvgdb}. We then use the definition of the virial radius {\blue} at redshift z=0 in the spherically symmetric halo:

\begin{equation}
R_\mathrm{vir}=\left(\frac{3 M_\mathrm{halo}}{4 \pi  \Delta_\mathrm{cr}  \rho_\mathrm{cr}}\right)^{\frac{1}{3}},
\end{equation}

\noindent
where $\mathrm{\rho_{cr}}$ is the critical mean density of the Universe, $\Delta_\mathrm{cr}$ is a factor that defines over-densities and $M_\mathrm{vir}$ is the halo mass. Common values for $\Delta_\mathrm{cr}$  range from 100 to 500 (or even higher) \citep[see Chapter 2.1 of][for more details about different values of $\Delta_\mathrm{cr}$]{Coe}.  If we assume that $\rho_\mathrm{cr}=137 \hspace{0.1cm} \mathrm{M_{\odot}/kpc}$ and $\Delta_\mathrm{cr}=100$, then we have:

\begin{equation}
R_\mathrm{vir}=259\left(\frac{M_\mathrm{halo}}{10^{12}\mathrm{M}_\odot}\right)^{1/3},
\label{Rvir}
\end{equation}

\begin{figure}[t]
\begin{center}

{\includegraphics[angle=0,height=8truecm,width=9truecm]{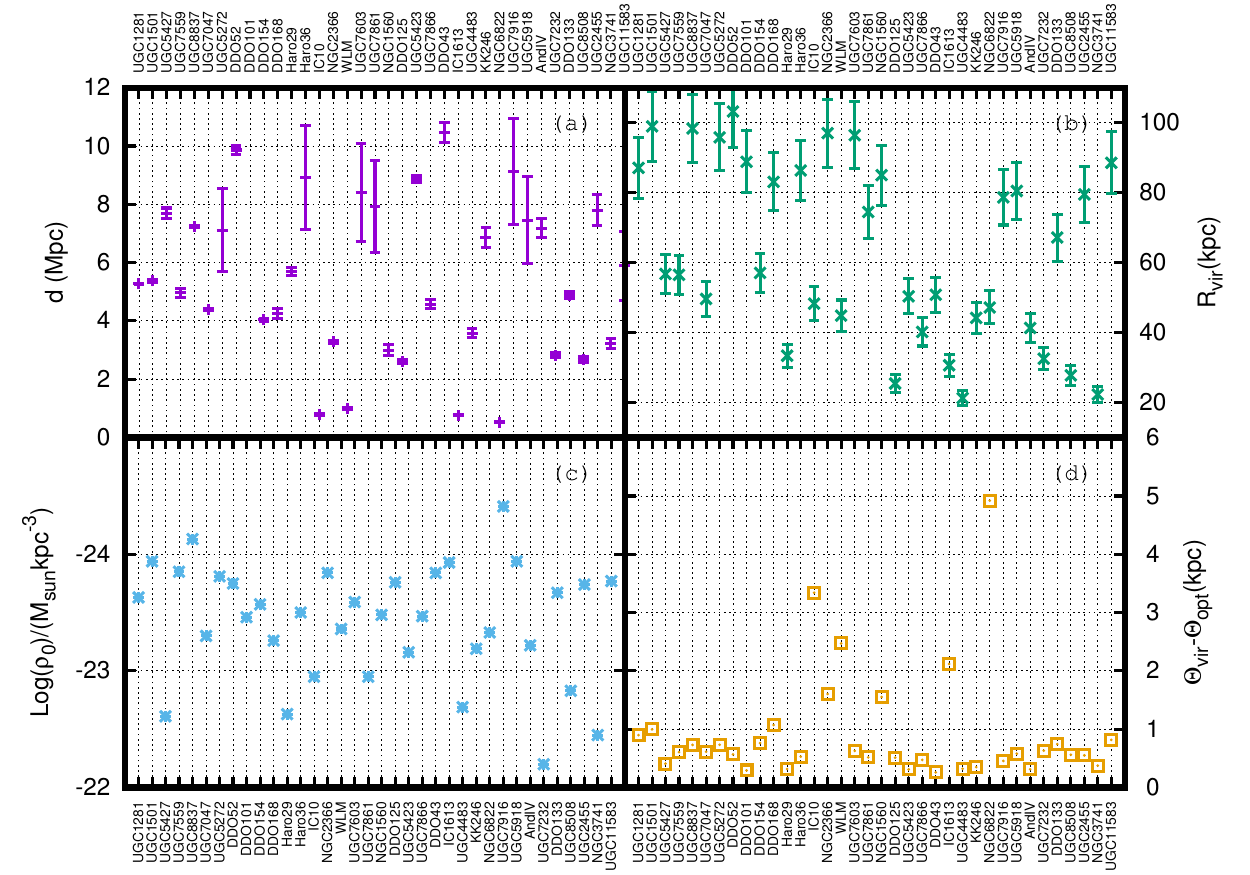}}

\caption {\centerlastline \footnotesize For each source we show the  parameters (reported also in Tables~\ref{Tab:Irr}~and~\ref{angles}) that are relevant for the calculation of the astrophysical factor. Panel (a): distance. Panel (b): virial radius. Panel (c): central density. Panel (d) $\theta_\mathrm{vir}-\theta_\mathrm{opt}$ (see text for details).  $\hspace{8cm}$}  
\label{Selection4P}
\end{center}
\end{figure}

\vspace{1cm}
\noindent
where the masses of the DM halos $M_\mathrm{halo}$ of galaxies in our sample are given in Table \ref{Tab:Irr}. In this table and in Fig. \ref{Selection4P} we also report the distance $d$, its error $\Delta d$, the optical radius $R_\mathrm{opt}$, structural parameters of the DM halo ($r_\mathrm{0}$, $\rho_\mathrm{0}$, $R_\mathrm{vir}$), the inclination, the stellar disk mass $M_\mathrm{disk}$ and the morphological~classification (MC). 

\begin{table*}
  \centering
  \footnotesize
  	\caption{Sample of 36 dwarf irregular galaxies.}
 \resizebox{\textwidth}{!}{  
  \begin{tabular}{|c|c|c|c|c|c|c|c|c|c|c|c|c|c|}

\hline
Name & $l$& $b$ & d & $\Delta$d   & $R_\mathrm{opt}$&   $ r_0 $	&	$\rho_0$  & $M_\mathrm{halo}$ & $R_\mathrm{vir}$ & incl&$M_\mathrm{disk}$ &MC&BCD\\
    & (deg) & (deg) & (Mpc) & (Mpc)  &  (kpc) & (kpc) & ($10^7\frac{M_\odot}{kpc^3}$)  & ($10^{10}M_\odot$) & (kpc) &  (deg) &  ($10^{8}M_\odot$) &---& boh \\
\hline	
UGC1281 & 136.9 & -28.7 &  5.27  & 0.02   & 3.39  &  3.20  & 3.5 & 3.7 & 86.0&90&1.4& Sdm&---\\
\hline
UGC1501 & 140.9 & -31 & 5.37  & 0.05  &		4.55   &	 4.80 &  1.7 & 5.4 & 97.9&75 &1.8&SBdm&---\\
\hline
UGC5427 & -160.6  & 53.4 &   7.69    &0.18   &	1.31    &	 0.86 &  36.4 & 1.0 & 55.8&55&0.5&Sdm&---\\
\hline
UGC7559 & 148.6 & 78.7 &  4.97  & 0.16   &	2.85    & 2.52 &   2.1 & 1.0 & 55.6&65&0.5&IBm&---\\
\hline
UGC8837 &103.7 & 60.8 &  7.24  &0.03   &	5.12  &	 5.65  &  1.1 & 5.3 & 97.3& 91.9 &1.7& IB(s)m&---\\
\hline
UGC7047 & 138.9 & 63.0 & 4.39    & 0.04  &	1.82&   	 1.36  &  7.5 & 0.7 & 48.7&44&0.3&IAm/BCD& 1\\
\hline
UGC5272 &-164.6	& 50.6  & 7.11    & 1.42  &	4.09&	 4.14 &  2.3  & 4.9 & 94.8&59&1.7&Im&--- \\
\hline
DDO52  & 179.1 	& 35.2  & 9.86  &  0.14 &	4.16  & 4.24 &   2.6 & 6.1 & 102.2&43&2.1&Im&---\\
\hline
DDO101 & -170.3 &	77.1  &16.60   & 3.32   &	3.10  & 2.71 &   5.1 & 3.9 & 87.8&51&1.5&Im&--- \\
\hline
DDO154 & 35.1&	89.4  & 4.04   & 0.06   &	2.40  &	 1.98  & 4.0 & 1.0 & 56.1&68.2&0.5&IB(s)m&--- \\
\hline
DDO168 &110.7&	70.7  & 4.25    & 0.16  &	2.59  & 	 2.20 &  8.2& 3.2 & 82.1 &46.5&1.3&IBm&---\\
\hline
Haro29  &134.1&	68.1 & 5.70 & 0.13   &	0.90   &	 0.51 &   34.2& 0.2& 32.7 &67&0.1&S/BCD&2\\
\hline
Haro36  &124.6&	65.5 & 8.91  & 1.78  &	3.11  &	 2.84 &   4.7 & 3.6 & 85.3&70&1.4&Im/BCD&3\\
\hline
IC10  & 119.0	 & -3.3   & 0.79  & 0.04 &	1.44   &  	 0.98  &  16.6 & 0.6 & 47.4&47&0.3&IBm/BCD&2\\
\hline
NGC2366  &146.4&	28.5 &  3.28  & 0.05 &	4.20  & 	 4.30 &  2.2 & 5.1 & 96.0 &68&1.7& IB(s)m&---\\
\hline
WLM  &75.9&	-73.6   &     0.98 & 0.03  &	1.76 &	 1.29  &  6.5 & 0.5 & 44.0&74&0.3& IB(s)m&---\\
\hline
UGC7603 &-108.4&	83.3&    8.40  & 1.68 &	3.56  &	 3.42 &   3.8& 5.0 & 95.4&78&1.8 &SB(s)d&---\\
\hline
UGC7861 &130.2&	75.7   & 7.91  & 1.58 &	1.98  &	 1.52  &  16.5 & 2.3 & 73.5&47&1.0 &SAB(rs)m&---\\
\hline
NGC1560  &138.4&	16.0  & 2.99  &0.19  &	3.04  &	 2.75  &  4.9& 3.4 & 84.0&82&1.3 &SA(s)d&---\\
\hline
DDO125  &137.7 &    73.0 &  2.61  &0.06  &	1.50 &	 1.04 &    2.6& 0.1 & 24.8&63&0.1&Im&---\\
\hline
UGC5423 &	140.0&	40.8&    8.87  &0.12  &	1.69  & 	 1.22    & 10.3 & 0.7 & 49.5 &56&0.4&Im/BCD &1\\
\hline
UGC7866 & 131.9	&78.5  & 4.57  	&0.15  &1.74 & 	 1.27  &  5.1 & 0.4 & 39.4&44&0.2&IAB(s)m&---\\
\hline
DDO43   &	177.8&	23.9  & 10.47 	&0.34  &2.62  &	 2.24  &  2.1 & 0.7 & 49.9&40.6&0.3&Im&---\\
\hline
IC1613  & 129.7	& -60.6 &  0.76   & 0.02 &	1.92 &  	 1.46  &  1.7 & 0.2 & 30.0&48&0.1&IB(s)m&---\\
\hline
UGC4483 &145.0&	34.4  &  3.58  	& 0.15 &0.67 &	 0.34 & 30.6  & 0.1 & 20.7 &58&0.04&Im/BCD&2\\
\hline
KK246*   &9.7&     -28.4  &  6.86 & 0.35    &	1.57 & 	 1.11   & 9.5  & 0.5 & 43.4&25 &0.3&Ir&---\\
\hline
NGC6822 &25.3	    &   -18.4&  0.52  &0.02   &	1.79  &	 1.32  &  7.0& 0.6& 46.3&58&0.3&IB(s)m&---\\
\hline
UGC7916 &134.2&	82.6 & 9.12 & 1.82	&5.22    &    	 5.80 &   0.6 & 2.7 & 77.6&74&1.0& Im&---\\
\hline
UGC5918 & 140.9	& 47.1 & 7.45 & 1.49 &	3.90   &	 3.88  &  1.7 & 2.9 & 79.5&46&1.1& Im&---\\
\hline
AndIV*  &121.1&	-22.3 &  7.18  & 0.33 &	1.52  &	 1.06 &   8.9& 0.4& 40.5&62&0.2&Ir&---\\
\hline
UGC7232 & 160.6	   &     77.6&  2.83 & 0.08  &	0.68 & 	 0.35 &   93.4 & 0.2 & 31.9&59&0.1&Im&---\\
\hline
DDO133 & 164.3&	84.0 & 4.88   &0.11  &	2.88    	 &2.55 &   3.2 & 1.7 & 66.2&43.4&0.7&Im&---\\
\hline
UGC8508 &111.1 &      61.3 & 2.67  & 0.10 & 0.89 & 0.51 & 22.0 &0.1&27.1&82.5&0.1&IAm&---\\
\hline
UGC2455 & 156.3    &  -29.2& 7.80 &0.54  & 3.4& 3.21& 2.7&2.8&78.5&51&1.1&IB(s)m&---\\
\hline
NGC3741 & 157.6  &    66.4 & 3.22&0.16  & 0.60& 0.29& 52.8 & 0.1&21.8&64&0.1&Im&---\\
\hline
UGC11583* & 095.6  &   12.3 & 5.89 & 1.18 & 3.75& 3.67& 2.5&3.8& 87.5&80&1.4&Ir&--- \\

\hline
\hline

\end{tabular}
}
\begin{justify}
Columns: (1) galaxy name; (2)-(3) source position in the sky in galactic coordinates, longitude $l$ and latitude $b$ respectively; (4)-(5) distance and the associated error; (6) optical radius;
(7) DM core radius; (8) DM central density; (9) DM halo mass; (10) virial radius; (11) inclination (an inclination of $90^\circ$ corresponds to an edge on galaxy, $0^\circ$ is a face on galaxy); (12) stellar disk mass; (13) morphological classification (MC) where the codes come from the Third Reference Catalog of Bright Galaxies (RC3) \citep{devaucouleurs}; (14)-BCD references. Here we also mark if a galaxy was identified as a BCD and its corresponding reference. Three of the galaxies  in the sample (with asterisks next to their names) are not present in the RC3 catalog, therefore we suggest their morphology following \citet{karachentsev}. The errors are estimated to be of $15\%$ on $\rho_0$, $r_0$, and $10\%$ on $R_\mathrm{vir}$ and $R_\mathrm{opt}$. 
\noindent
{\bf BCD references}: 1-\citet{parodi}, 2-\citet{gildepaz}, 3-\citet{thuan}
\end{justify}

\label{Tab:Irr}

\end{table*}

 \section{Gamma-ray telescopes}
 \label{telescopessection}

\begin{figure}[t]
\begin{center}
{\includegraphics[angle=0,height=8truecm,width=9truecm]{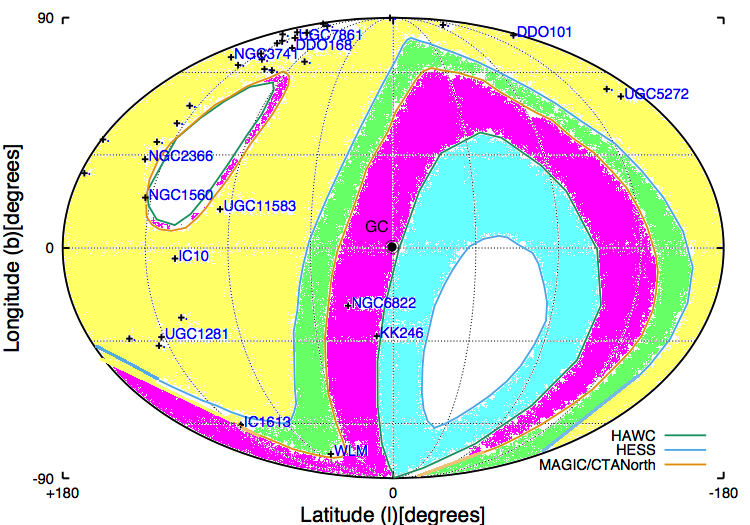}}
\raggedleft 
\caption { \centerlastline \footnotesize Sky map: position of the 36 dIrr galaxies in the sample and sky coverage of several gamma-ray telescopes. Overlapping regions are showed in different colours. We stress that Fermi-LAT covers almost the whole sky. The yellow region is covered by MAGIC, CTA-north and HAWC. The green region can be observed by the previous and HESS. The violet region is covered by the previous experiments except MAGIC and CTA north. The cyan region can be observed only by HESS so far. CTA-south will also cover this region in the next future. The black point indicates the Galactic Center (GC) region. Among other galaxies of the sample, we show here the position of NGC6822, IC10 and WLM. $\hspace{10cm}$}
\label{skymap}
\end{center}
\end{figure}

\begin{table}
\centering
\begin{center}
\resizebox{9cm}{!}{
\begin{tabular}{|c|c|c|c|c|}
\hline
Experiment& Fermi-LAT & HESS I (II) & HAWC & CTA North/South\\
\hline
E range & 20 MeV - 300 GeV & 0.03- 100 TeV & 0.1-100 TeV& 0.02-200 TeV \\
\hline
$\Delta$ E & $10\%$  & $15\%$ & $50\%$ & $10\%$\\
\hline
FoV (deg) & $>50$ &  5 (3.2) & wide & 10 \\
\hline
$\theta_\mathrm{PSF}$ (deg) & $0.1-0.5$  & $0.1$ & $0.1$& $0.05$ \\
\hline
$\mathrm{A}_\mathrm{eff} (\mathrm{cm}^2)$& $10^4$ & $1(6)\times10^6$ &$10^5$ & $10^{10}$ \\
\hline
$t_\mathrm{exp}$ & $10$yr   &  100 h & 5 yr &1000 h\\
\hline
\hline
\end{tabular}
}
\end{center}
\caption{\centerlastline \footnotesize {Energy range, energy resolution ($\Delta$E), field of view (FoV), angular resolution $\theta_\mathrm{PSF}$, effective area $A_\mathrm{Eff}$ and the expected observation time $t_\mathrm{exp}$ for the Fermi-LAT satellite, the air and water Cherenkov observatories HESS, HAWC \citep{HAWC,HAWC1,HAWC2} and the future CTA telescopes, respectively. The angular resolution in Fermi-LAT strongly depends on the energy.  $\hspace{10cm}$}}

\label{telescopes}
\end{table}

Detectors on satellites as well as ground-based air and water Cherenkov arrays are dedicated to the detection and study of gamma-ray sources. Different instruments cover different regions of the sky, energy range, resolution angle and field-of-view. We consider a selection of gamma-ray telescopes and we report their main characteristics in Table~\ref{telescopes}. In Fig. \ref{skymap} we show the position of our sample of galaxies (as given in Table \ref{Irr}) with respect to the sky coverage of several instruments. Both the Fermi Large Area Telescope (Fermi-LAT) and the combination of the Cherenkov Telescope Array (CTA) South and North are expected to cover the whole sky. The Major Atmospheric Gamma Imaging Cherenkov Telescopes (MAGIC), The High Energy Stereoscopic System (HESS) and the High-Altitude Water Cherenkov Observatory (HAWC) instruments have limited sky coverage. Most of the galaxies of our sample are in the sky region covered by Fermi-LAT, HAWC, MAGIC and future CTA-North telescope. A minority of them are also observable with the HESS or the future CTA-South telescope, but not by MAGIC (see Fig.~\ref{skymap}).

\section{Gamma-ray flux from dark matter annihilation}
\label{gammarayDMsection}

Gamma-ray astronomy represents one of the most promising methods to search for the DM indirectly \citep{GCCR4,GCCR,GCCR1,GCCR3}.
Since the properties of large scale structures observed in our Universe imply that DM is fairly cold and its particles are relatively slow moving, the Weakly Interacting Massive Particles (WIMPs) are one plausible DM candidates so far \citep{BUB,FMW}. In the generic WIMP scenario, two massive DM particles can annihilate producing two lighter SM particles. In this framework, the differential gamma-ray flux from two annihilating DM particles in galactic sources is:

\begin{equation}
\frac{\mathrm{d}\,\phi_{\gamma}^{\mathrm{DM}}}{\mathrm{d}\,E_{\gamma}} =
\frac{dP}{dE}\, <J>_{\Delta\Omega}.
\label{flux}
\end{equation}

\noindent
The first term on the r.h.s. is the particle physics dependent part:

\begin{equation}
\frac{dP}{dE}=\frac{1}{8\pi \mathrm{m}_\mathrm{DM}^2}\sum_i\langle\sigma_i v\rangle
\frac{\mathrm{d}\,N_\gamma^i(\mathrm{m}_\mathrm{DM})}{\mathrm{d}\,E_{\gamma}}\, ,
\label{Eq:P}
\end{equation}

\noindent
that depends on the DM particle mass $\mathrm{m}_\mathrm{DM}$ and the averaged
annihilation cross-section $\langle\sigma_i v\rangle$ of two DM particles into two SM particles (labeled by the subindex $i$). After the main DM annihilation event, the chain of subsequent hadronization and decay events of SM particles, such as quarks and leptons, produce secondary fluxes of cosmic rays, such as gamma rays, neutrinos, antimatter etc. Due to the non-perturbative quantum chromodynamic effects, the analytical
calculation of these decay chains is a hard task to be accomplished
and therefore it requires Monte Carlo events generators
such as PYTHIA \citep{PYTHIA} or HERWIG \citep{HERWIG} particle physics software. Here, we use Cirelli's code \citep{Cirelli} and we include electro weak corrections \citep{EW}. The uncertainty related to the choice of the Monte Carlo events generator software that produces the simulated gamma-ray flux was studied, among others, in \cite{MC}. We assume here that the DM particle is a Majorana fermion. An extra factor of  1/2
would appear in Eq. \ref{Eq:P} in the case of symmetric Dirac fermion dark matter. \\

The second term on the r.h.s. of Eq. \ref{flux} represents the astrophysical factor:
\begin{equation}
<J>_{\Delta\Omega}=\frac{1}{\Delta\Omega}\int_{\Delta\Omega}\mathrm{d}\Omega\int_{l.o.s.} \rho^2(s) \mathrm{d}s
\label{Eq:J}
\end{equation}

\noindent
 i.e, the integral of the DM mass density
profile, $\rho(r)$, along the path (line of sight, $l.o.s.$) between 
the gamma-ray detector and the source divided by the solid angle $\Delta\Omega=2\pi(1-\cos{\theta})$. The allowed range of values for $\theta$ in the framework of point-like analysis is given by the Point Spread Function (PSF) of the instrument, as we will discuss in the following.

\begin{figure}[t!]
\begin{center}
{\includegraphics[angle=0,height=8truecm,width=9truecm]{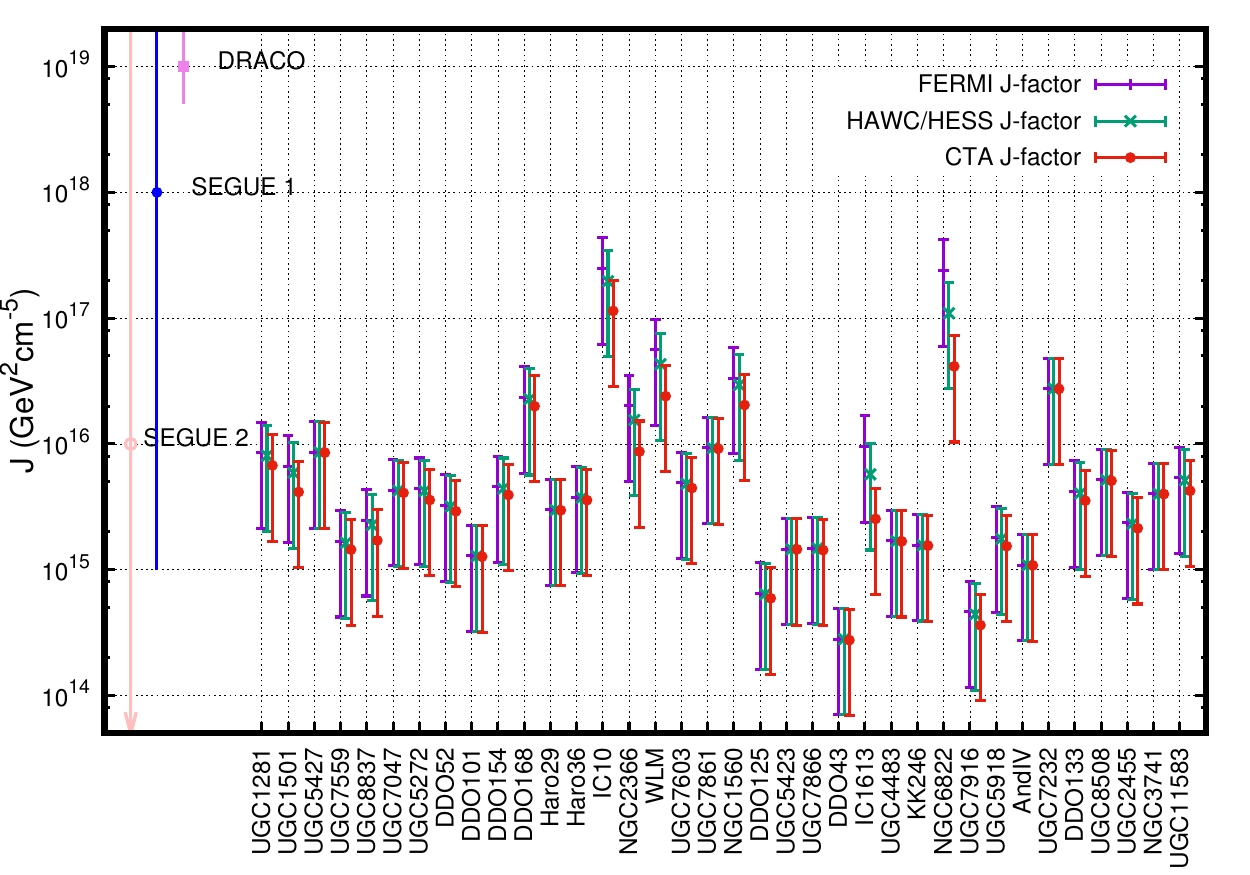}}
\caption {\centerlastline \footnotesize J-factors $J=\langle J\rangle_{\Delta\Omega}\Delta\Omega$ for the angular resolution (PSF) of different telescopes assuming the point-like source analysis. We use the following PSF: $\theta_\mathrm{Fermi-LAT}=0.5^\circ$ (purple points), $\theta_\mathrm{HAWC/HESS}=0.1^\circ$ (green points) and $\theta_\mathrm{CTA}=0.05^\circ$ (red points). The point-like analysis account for more than $95\%$ of the DM halo. The error bars account for the uncertainties in the DM density distribution profile for the estimation of the J-factor. For the comparison, we also show the astrophysical factors and the associated error bars of three dSphs: SEGUE1, SEGUE2 and DRACO \citep{triaxSph}.$\hspace{3cm}$
}
\label{thetaEXPJ}
\end{center}
\end{figure}

\begin{figure}[t]
\begin{center}
{\includegraphics[angle=0,height=7.7truecm,width=9truecm]{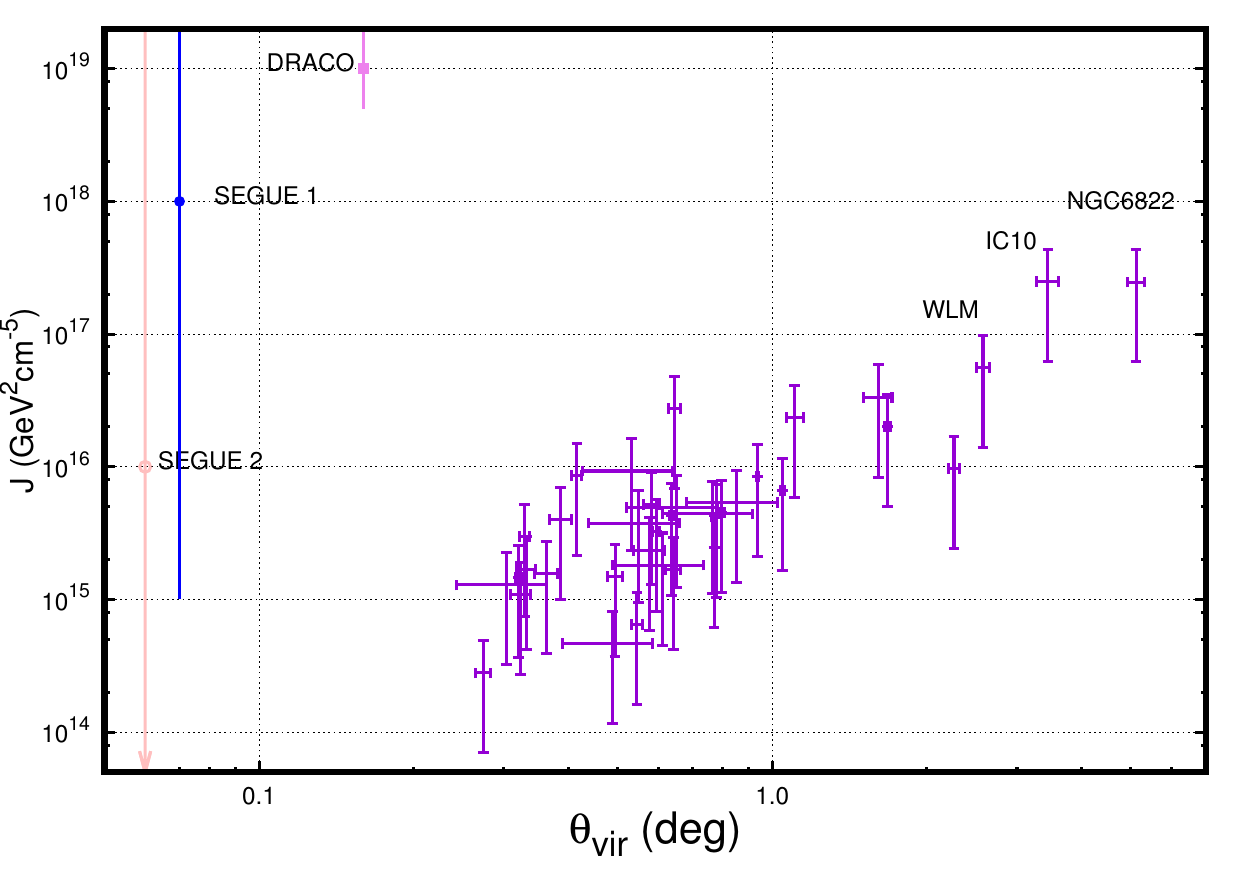}}
\caption {\centerlastline \footnotesize{ $J_\mathrm{vir}$-factor as a function of the angle $\theta_\mathrm{vir}$, where $\theta_\mathrm{vir}$ is the projection of the virial radius on the sky. The error bars represent the $75\%$ of} $J_\mathrm{vir}$. Unlike the point-like hypothesis in Fig. \ref{thetaEXPJ}, in this case we also have errors on $\theta_\mathrm{vir}$ corresponding to the uncertainty on the estimation of virial radio $R_\mathrm{vir}$ and the distance of the source target $d$ (see Tables \ref{Tab:Irr} and \ref{angles} and text for more details. For the comparison, we also report the astrophysical factors and the associated error bars of three dSphs: SEGUE1, SEGUE2 and DRACO \citep{triaxSph}. The angular dimensions of the latter are given by the corresponding tidal radii.\hspace{10cm} }
\label{thetavirJ}
\end{center}
\end{figure}

\begin{figure}[t]
\begin{center}
{\includegraphics[angle=0,height=7.3truecm,width=9truecm]{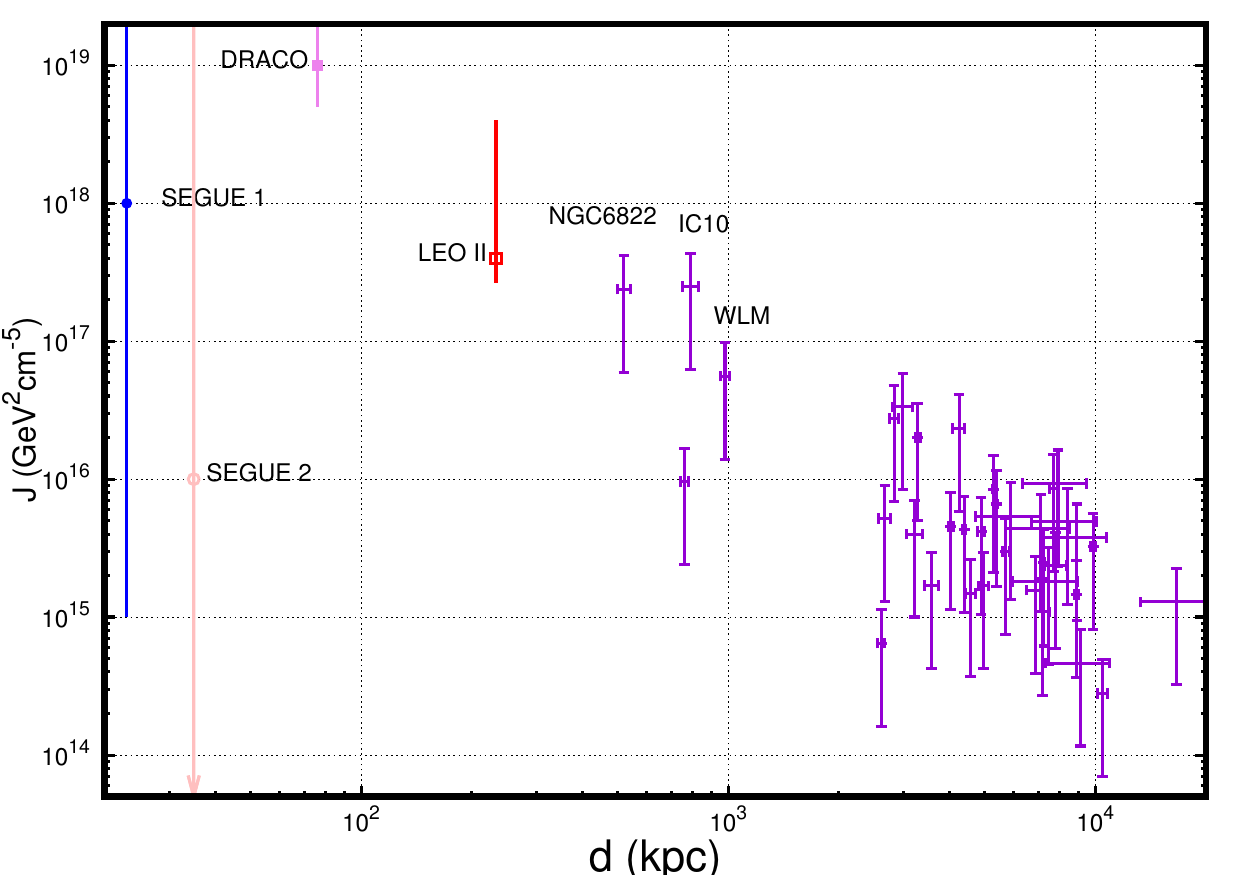}}
\caption {\centerlastline \footnotesize{ $J_{0.5^\circ}$-factor as a function of the distance to the target. The J-factors of closer objects are promising with respect to some dSphs.\hspace{0.9cm} }}
\label{Jvsd}
\end{center}
\end{figure}

\begin{table*}[ht!]
\centering
\begin{center}
 \resizebox{\textwidth}{!}{ 
\begin{tabular}{|c|c|c|c|c|c|c|c|c|}
\hline
Name & $\theta_\mathrm{opt}$ (deg)    &   $\theta_\mathrm{vir}$ (deg) & $\theta_\mathrm{vir}-\theta_\mathrm{opt}$ &$J_{0.05^\circ}$ ($\mathrm{GeV}^2\mathrm{cm}^{-5}$) &$J_{0.1^\circ}$ ($\mathrm{GeV}^2\mathrm{cm}^{-5}$) & $J_\mathrm{0.5^\circ}$ ($\mathrm{GeV}^2\mathrm{cm}^{-5}$) & $J_\mathrm{vir}$ ($\mathrm{GeV}^2\mathrm{cm}^{-5}$)   &	\\
\hline
UGC1281 	&$0.037\pm 0.004$ 	&$0.936\pm 0.003$ &$ 0.90\pm 0.01$ 	 & $6.74\times10^{15}$ & $8.08\times10^{15}$	& $8.46\times10^{15}$ & $8.46\times10^{15}$  & 			\\
\hline
UGC1501 	&$0.048\pm0.005$ 	&$1.04\pm0.01$ & $1.00\pm0.01$		  & $4.14\times10^{15}$	& $5.88\times10^{15}$	& $6.60\times10^{15}$ &$6.60\times10^{15}$ 	& 		\\
\hline
UGC5427	         &<0.01&	$0.42\pm0.01$ & $0.41\pm0.01$      & $8.52\times10^{15}$   & $8.56\times10^{15}$     &$8.56\times10^{15}$ & $8.56\times10^{15}$  	&	\\
\hline
UGC7559  	&$0.033\pm0.003$ 	&$0.64\pm0.02$& $0.61\pm0.02$		  & $1.44\times10^{15}$	& $1.63\times10^{15}$&$1.68\times10^{15}$ & $1.68\times10^{15}$ 		&	\\
\hline
UGC8837	         &$0.040\pm0.004$ 	& $0.771\pm0.003$ &$0.73\pm0.01$	           & $1.71\times10^{15}$  & $2.27\times10^{15}$      &$2.47\times10^{15}$ &$2.47\times10^{15}$ 	&		\\
\hline
UGC7047	         &$0.023\pm0.002$ &	$0.64\pm0.01$& $0.61\pm0.01$          & $4.08\times10^{15}$  &  $4.25\times10^{15}$  &$4.29\times10^{15}$ & $4.29\times10^{15}$ 		&	\\
\hline	
UGC5272	         &$0.03\pm0.01$ &	$0.7\pm0.2$& $0.7\pm0.2$		           & $3.58\times10^{15}$ & $4.23\times10^{15}$  &$4.41\times10^{15}$ & $4.42\times10^{15}$& \\			
\hline
DDO52	         &$0.024\pm0.002$ &	$0.59\pm0.01$& $0.57\pm0.01$	   & $2.91\times10^{15}$   &   $3.15\times10^{15}$   &$3.25\times10^{15}$ & $3.25\times10^{15}$ &\\
\hline	
DDO101	         &<0.01 	&$0.30\pm0.06$& $0.29\pm0.06$				  & $1.27\times10^{15}$  &  $1.29\times10^{15}$   &$1.29\times10^{15}$ &$1.29\times10^{15}$ 		&		\\
\hline
DDO154	         &$0.034\pm0.003$	&$0.8\pm0.1$& $0.8\pm0.1$			   & $3.92\times10^{15}$  &  $4.42\times10^{15}$  &$4.54\times10^{15}$ &$4.54\times10^{15}$ 		&		\\
\hline
DDO168	         &$0.035\pm0.004$ &$1.11\pm0.04$ & $1.07\pm0.04$		   & $1.99\times10^{16}$  & $2.27\times10^{16}$   &$2.34\times10^{16}$ & $2.34\times10^{16}$ 		& x		\\
\hline
Haro29      	&<0.01 &	$0.33\pm0.01$&$0.32\pm0.01$		   & $2.97\times10^{15}$	 & $2.98\times10^{15}$ &$2.99\times10^{15}$ &$2.99\times10^{15}$ 		&		\\
\hline
Haro36	         &$0.020\pm0.004$ 	&$0.5\pm0.1$& $0.5\pm0.1$			   & $3.57\times10^{15}$  &  $3.73\times10^{15}$    &$3.77\times10^{15}$ & $3.77\times10^{15}$ 		&		\\
\hline
IC10	                  &$0.10\pm0.1$&	$3.4\pm0.2$ & $3.3\pm0.2$			   & $1.14\times10^{17}$   &   $1.97\times10^{17}$     &$2.48\times10^{17}$ & $2.49\times10^{17}$ 	& x	\\
\hline
NGC2366        &$0.07\pm0.01$		&$1.68\pm0.03$ & $1.60\pm0.03$		   & $8.68\times10^{15}$  & $1.55\times10^{16}$ &$2.01\times10^{16}$ & $2.02\times10^{16}$ 	& x		\\
\hline
WLM	          &$0.10\pm0.01$ &$2.6\pm0.1$ & $2.5\pm0.1$				   & $2.40\times10^{16}$ & $4.30\times10^{16}$ & $5.59\times10^{16}$ & $5.62\times10^{16}$ 	& x			\\
\hline
UGC7603    &	$0.024\pm0.005$		&$0.6\pm0.1$& $0.6\pm0.1$		  & $4.52\times10^{15}$& $4.82\times10^{15}$	&$4.90\times10^{15}$ & $4.90\times10^{15}$ 			& \\
\hline
UGC7861    &	<0.01 &	$0.5\pm0.1$&$0.5\pm0.1$					   & $9.16\times10^{15}$ & $9.27\times10^{15}$ &	$9.30\times10^{15}$ &$9.30\times10^{15}$ 	& 	\\
\hline
NGC1560	&$0.06\pm0.01$	&$1.6\pm0.1$&$1.5\pm0.1$			           & $2.03\times10^{16}$& $2.95\times10^{16}$  &   $3.34\times10^{16}$ &$3.34\times10^{16}$ & x	\\
\hline
DDO125	&$0.033\pm0.003$	&	$0.54\pm0.01$&$0.51\pm0.02$			  & $5.91\times10^{14}$& $6.38\times10^{14}$   &  $6.48\times10^{14}$ &$6.48\times10^{14}$ 		&		\\
\hline 
UGC5423	&<0.01	&	$0.320\pm0.004$& $0.310\pm0.005$			   & $1.45\times10^{15}$	& $1.46\times10^{15}$  &   $1.46\times10^{15}$ &$1.46\times10^{15}$ 	&		\\
\hline
UGC7866	&$0.022\pm0.002$	&	$0.49\pm0.02$& $0.5\pm0.2$			   & $1.43\times10^{15}$&  $1.45\times10^{15}$  &   $1.49\times10^{15}$ &$1.49\times10^{15}$ 			&	\\
\hline
DDO43	&<0.01	&	$0.27\pm0.01$& $0.3\pm0.1$					   &$2.75\times10^{14}$	&  $2.80\times10^{14}$   &  $2.80\times10^{14}$ & $2.80\times10^{14}$ 	&		\\
\hline
IC1613	&$0.14\pm0.02$     & $2.26\pm0.06$&$2.11\pm0.07$			   & $2.53\times10^{15}$& $2.73\times10^{15}$  &    $9.57\times10^{15}$ & $9.70\times10^{15}$ 		& 		\\
\hline
UGC4483	&<0.01	&	$0.33\pm0.01$& $0.32\pm0.01$				   & $1.68\times10^{15}$	& $1.69\times10^{15}$  &    $1.69\times10^{15}$ &$1.69\times10^{15}$ 	&	\\
\hline
KK246	&<0.01      &	$0.36\pm0.02$& $0.35\pm0.02$				   & $1.55\times10^{15}$& $1.56\times10^{15}$  &    $1.57\times10^{15}$ &$1.57\times10^{15}$ 	&		\\
\hline
NGC6822	&$0.20\pm0.02$	&	$5.1\pm0.2$ &$4.9\pm0.2$			   & $4.13\times10^{16}$& $1.09\times10^{17}$  &    $2.39\times10^{17}$ &$2.46\times10^{17}$ & x	\\
\hline
UGC7916	&$0.03\pm0.01$     &$0.5\pm0.1$&$0.4\pm0.1$				   & $3.61\times10^{14}$& $4.41\times10^{14}$  &    $4.64\times10^{14}$&$4.64\times10^{14}$ & \\		
\hline
UGC5918	&$0.03\pm0.01$      	&$0.6\pm0.1$&$0.5\pm0.1$			           & $1.53\times10^{15}$& $1.75\times10^{15}$  &    $1.81\times10^{15}$ & $1.81\times10^{15}$ & \\		
\hline
AndIV	&<0.01    &	$0.32\pm0.1$& $0.31\pm0.02$					  & $1.08\times10^{15}$& $1.09\times10^{15}$ &     $1.09\times10^{15}$ & $1.09\times10^{15}$ 	& \\
\hline
UGC7232	&<0.01     &		$0.64\pm0.02$& $0.63\pm0.02$			  & $2.73\times10^{16}$& $2.75\times10^{16}$&$2.75\times10^{16}$ & $2.75\times10^{16}$ & x\\		
\hline
DDO133	&$0.034\pm0.003$     &	$0.78\pm0.01$& $0.74\pm0.02$		 & $3.53\times10^{15}$& $4.05\times10^{15}$  &    $4.18\times10^{15}$ &$4.18\times10^{15}$ &	\\
\hline
UGC8508& $0.020\pm0.002$& $0.58\pm0.02$&$0.56\pm0.02$ 			 & $5.09\times10^{15}$&  $5.16\times10^{15}$   &   $5.17\times10^{15}$ &$5.17\times10^{15}$ & \\		
\hline
UGC2455 &$ 0.025\pm0.003$  & $0.58\pm0.04$&$0.55\pm0.04$			 & $2.13\times10^{15}$&  $2.32\times10^{15}$ &    $2.36\times10^{15}$ &$2.36\times10^{15}$ 	&\\
\hline
NGC3741 & <0.01&  $0.38\pm0.02$&$0.38\pm0.02$					 & $3.99\times10^{15}$& $4.00\times10^{15}$  &    $4.00\times10^{15}$ &$4.00\times10^{15}$ 			&	\\
\hline
UGC11583 & $0.04\pm0.01$ & $0.8\pm0.2$&$0.8\pm0.2$				 & $4.24\times10^{15}$& $5.14\times10^{15}$ &   $5.40\times10^{15}$ &$5.40\times10^{15}$  			 &	\\
\hline
\hline
\end{tabular}
}
\end{center}
\caption{\centerlastline \footnotesize{Projected angles and the associated uncertainties correspond to the optical ($\theta_\mathrm{opt}$) and the virial ($\theta_\mathrm{vir}$) radii for each dIrr galaxy (see text for details). Astrophysical J-factors calculated as $J=\langle J \rangle_{\Delta\Omega}\Delta\Omega $, where $\langle J \rangle_{\Delta\Omega}$ is presented in Eq. (\ref{Eq:J}). We perform this calculation for several angular resolution associated with different experiments: $0.05^\circ$ for the next CTA, $0.1^\circ$ for HESS, HAWC and Fermi-LAT higher energies, $0.5^\circ$ Fermi-LAT. The J-factors ideally calculated on the virial angular dimensions are also shown for comparison (details are given in Appendix~\ref{AppB}). dIrrs marked with "x" are included in the combined analysis of 7 best galaxies in the sample. \hspace{15cm}
}
}
\label{angles}
\end{table*}

\section{The astrophysical factor}
\label{Jfactor}

In this section we will focus on the astrophysical J-factor of the galaxies in the sample. The J-factor depends on the DM density distribution profile, the angular resolution of the telescope and the distance to the target. Instead, as we will discuss in Section~\ref{gammarayBg}, given a thermal WIMP candidate, the particle physics dependent part $P(\mathrm{m}_\mathrm{DM},\langle\sigma_i v\rangle)$ introduced in Eq. (\ref{flux}), is totally independent of the astrophysical context. The DM density distribution parameters for our sample of 36 galaxies are listed in Table \ref{Tab:Irr} and shown on Fig.~\ref{Selection4P}, while the angular resolution of different telescopes are summarized in Table \ref{telescopes}. The interested reader can find the details on the calculation of the astrophysical J-factor in the Appendix \ref{AppB}. The J-factor calculation gives a first order estimation on the  competitiveness of dIrrs with respect to dSphs for DM searches. In the following we present two different approximations in order to calculate the J-factor in our sample, that is (i) fixing the telescope PSF and (ii) ideally accounting for the whole dIrr virial radius.\\

(i) \textit{Point-like J-factors} - Following  Eq. (\ref{Eq:J}) we calculate the J-factors for the sample of 36 dIrr galaxies in light of a point-like source analysis, where the angular resolution is set by the PSF of the instrument. In details, we assume different values of the solid angles (or integration angles) $\Delta\Omega$ that correspond to the angular resolutions  (or PSF) of various gamma-ray telescopes: $\theta=0.05^\circ, 0.1^\circ, 0.5^\circ$, that are CTA, HESS/HAWC and Fermi-LAT, respectively. 
The J-factors and their uncertainties are shown in Fig. \ref{thetaEXPJ}. Here the error bars represent the uncertainties of the DM density profile. Then, the $15\%$ error on the DM density distribution parameters $\rho_0$ and $r_0$ introduces an uncertainty of $20\%-60\%$ on the density distribution itself, that is $75\%$ of the astrophysical J-factor. We neglect the uncertainties on both the extreme limits of integration along the $l. o. s.$ and the solid angle since these contributions are expected to be negligible. Let us notice that the J-factor of NGC6822 decreases of almost one order of magnitude ($83\%$) decreasing the integration angle form $0.5^\circ$ (Fermi-LAT) to $0.05^\circ$ (CTA). We also notice that for the most galaxies of our sample the J-factors calculated for the PSF of $0.5^\circ$ are identical to that calculated using their virial dimensions (see the discussion in the next paragraph and Table \ref{angles}).

(ii) \textit{Virial J-factors} - In Eq. (\ref{Eq:J}) the solid angle $\mathrm{\Delta\Omega}$, instead of being fixed by the PSF of the instrument, is varying with the dimension of the source.
The virial J-factors are listed in Table~\ref{angles} and plotted as a function of the projected angular dimensions $\theta_\mathrm{vir}$ in Fig. \ref{thetavirJ}. This angle depends on the virial radius and the distance to the source as: $R_\mathrm{vir}^2=d^2\sin^2{\theta_\mathrm{vir}}$ (see Appendix \ref{AppB} for details). In Fig. \ref{thetavirJ} it is evident that the virial J-factor increases with the angle $\theta_\mathrm{vir}$. However, comparing Fig.~\ref{thetaEXPJ} with Fig. \ref{thetavirJ}, it becomes clear that taking into account the full DM halo up to $R_\mathrm{vir}$ or just the radius corresponding to the PSF of an instrument, does not affect much the resulting value of the J-factor. Let us notice that the J-factor of NGC6822 increases of only $4\%$ when we move from the integration angle of $0.5^\circ$ to $5^\circ$. The latter corresponds to the angular dimension of the virial radius. The uncertainties on the virial J-factors in Fig. \ref{thetavirJ} are calculated as for the point-like analysis. Instead, the error on $\theta_\mathrm{vir}$ is obtained by taking into account that the value of the virial radius $R_\mathrm{vir}$ in galaxies is independent of the distance to them. Therefore, we can calculate the maximum error on $\theta_\mathrm{vir}$ for each galaxy as:

\begin{equation}
\Delta\theta_\mathrm{vir}=\left(\sum_{x_n=d,R_\mathrm{vir}}\left(\frac{\partial\theta_\mathrm{vir}(x_n)}{\partial x_n}\bigg|_{x_n}\Delta x_n\right)^2\right)^{1/2}.
\end{equation}

\noindent
The error bars on the $\theta_\mathrm{vir}$ are listed in Table \ref{angles} and shown~on~Fig. \ref{thetavirJ}.  \\

In both cases of the point-like and the virial J-factors, the highest values
are obtained for NGC6822, IC10, WLM, that are the closest objects in the sample, with distances of 520 kpc, 790, kpc and 980 kpc respectively (Panel (a) of Fig.~\ref{Selection4P}), central densities of $10^7-10^8 M_\odot$  (Panel (c) of Fig.~\ref{Selection4P}) and halo mass of $\sim 10^{10} M_\odot$ (see Table \ref{Tab:Irr}). In fact, the PSF of Fermi-LAT ($\theta \approx 0.5^\circ$) corresponds to a radius of $\sim$~4.5~kpc, $\sim$6.9 kpc and $\sim$8.5 kpc for the NGC6822, IC10 and WLM galaxy respectively. Let us notice that the average total DM halo mass in the spatial region of dIrrs enclosed within the Fermi-LAT PSF is $\sim 10^9 M_\odot$. 
This mass is two orders of magnitude bigger than the mass enclosed in the tidal radius of a typical dSph galaxy ($\sim10^7M_\odot$). \\

Finally, on Fig. \ref{Jvsd} we plot the J-factors for a PSF of $0.5^\circ$ versus the distance to the dIrrs of our sample and several dSphs. It appears clearly that dIrr galaxies although staying at higher distances, in some cases having similar values of the J-factors to that of some dSphs. It is also clear that there is some kind of dependence of the value of the J-factor on the distance to the object. This fact has been already addressed previously for the Milky Way satellites (see e.g. \citep{Fermi-LAT:2016uux}). Therefore, it is important to notice that there are certainly more dIrrs galaxies at the distance less than $\sim$~2~ Mpc than presented here. However, not all of them have been detected yet and not all the detected dIrrs have available kinematics so far (see the discussion section).

\subsection{Baryon and dark matter spatial distribution} 
 
 \vspace{0.3cm}
 
In this section we will discuss the \textit{extra} astrophysical contribution to the gamma-ray flux expected from the DM annihilation events in dIrr galaxies. We further assume that this \textit{extra} astrophysical background is enclosed in the corresponding star-forming region. The latter is associated with their optical radii, that are several times smaller than the corresponding virial radii.

\begin{figure}[t]
\begin{center}
{\includegraphics[angle=0,height=7truecm,width=9truecm]{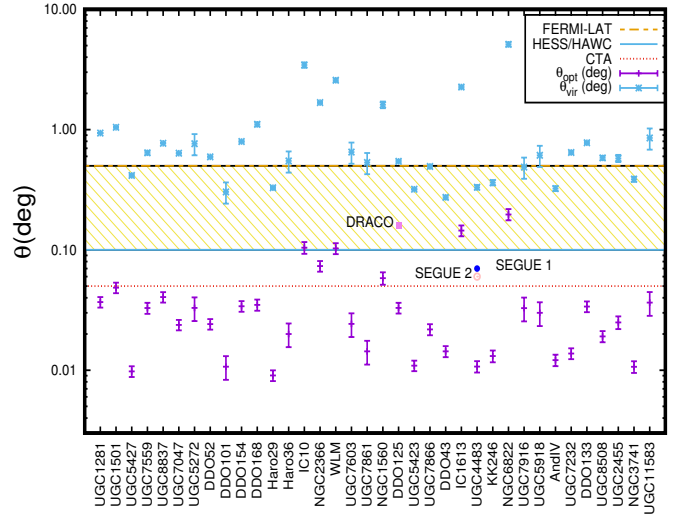}}
\caption {\centerlastline \footnotesize {Angular dimensions of different sources and resolutions angles of several gamma-ray experiments. For each source, we show: the projected virial (blue squares) and optical (purple points) radii. Angular resolutions are given for: Fermi-LAT ($0.5^\circ$) (yellow-black dashed line), HESS and HAWC ($0.1^\circ$) (blue solid line), and CTA ($0.05^\circ$) (red dotted line). The angular resolution of Fermi-LAT strongly depends on the energy (yellow diagonally cross-hatched region). The star-forming (optical) regions of our sample of dIrr galaxies are unresolved by Fermi-LAT with $0.5^\circ$ of the resolution power. For comparison, we also show the angular dimensions (the projected tidal radii) of three dSph galaxies: DRACO, SEGUE1 and SEGUE2.  $\hspace{10cm}$
}}
\label{angularres}
\end{center}
\end{figure}

A comparison of the optical and the virial projected angles  of dIrr galaxies on the sky (or angular dimensions)
with the PSF of several instruments is given in Table~\ref{angles} and Fig. \ref{angularres}. This figure
illustrates that the star-forming regions of most of our galaxies are not resolved and they may contribute to the diffuse Isotropic Gamma-Ray Background (IGRB) observed by Fermi-LAT \citep{Fornasa, Fornasa1,Linden:2016fdd,DiMauro:2015ika}, that is the residual gamma-ray emission after subtraction of the emission from resolved sources and the Galactic diffuse foreground induced by cosmic rays. Four galaxies of our sample (NGC6822, IC10, WLM and IC1613) have optical regions of $\sim0.1^\circ-0.2^\circ$, thus they could be potentially resolved by several devices. However, in the next section we will show that they still remain undetected. In fact, their estimated star-forming gamma-ray luminosity is lower than the Fermi-LAT point-like detection threshold. 
Most of the 36 galaxies has optical angular dimension smaller than $\sim0.05^\circ$, staying unresolved also for the next-generation of gamma-ray telescopes. On the contrary, their DM virial halos are bigger than $\gsim0.2^\circ$, and could potentially appear as marginally-extended sources  (see Table \ref{angles} and Fig. \ref{angularres}). However, also in the case of any detection, distinguishing between a point-like or marginally extended source would not be possible, because of the convolution with the PSF of the instrument. For this reason, we need a model for the \textit{extra} astrophysical background component in dIrrs.  We will show that this component is negligible with respect to both the gamma-ray flux expected from DM annihilation events and the diffuse and isotropic gamma-ray background components adopted in the analysis of \textit{classical} dSph galaxies. This fact allows to set constraints on the DM particle via the point-like analysis of dIrr galaxies.

\section{The gamma-ray emission}
\label{gammarayBg}

As previously introduced, in order to search for the secondary gamma-ray DM signature in dIrrs, we need to take into account the \textit{extra} gamma-ray contamination, associated with their star-forming regions. Notice, that this gamma-ray background we call \textit{extra} because it is absent or considered negligible in \textit{classical} dSph galaxies.
Even if the difference in the distribution of baryons and DM in these objects may favour the DM detection in the extended halo with respect to the unresolved star-forming region, the mask can not be applied due to the limited PSF of the current generation of gamma-ray telescopes. Therefore, identifying and modelling the possible astrophysical gamma-ray emission in 
dIrrs is essential before searching for the DM annihilation signal.

\begin{figure}
\includegraphics[angle=0,height=7.0truecm,width=9.0truecm]{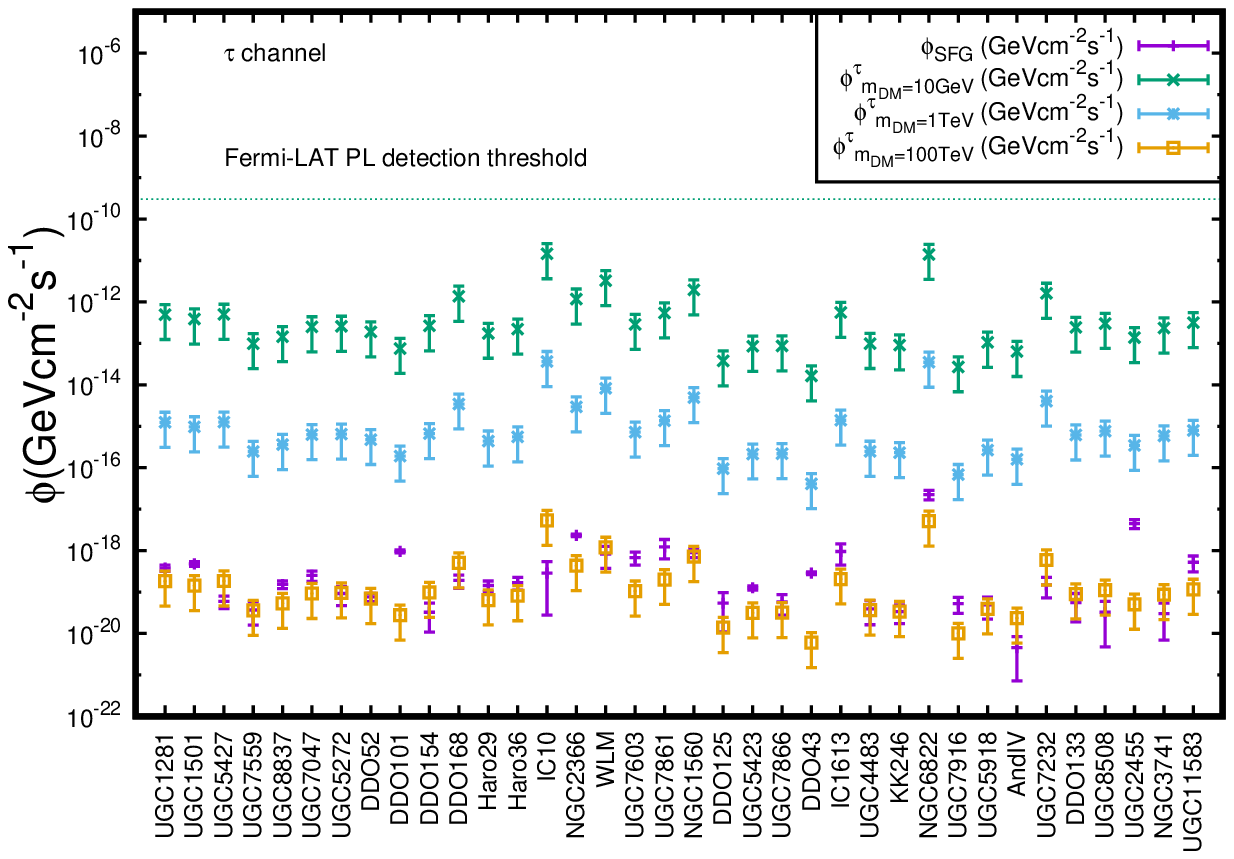}
\includegraphics[angle=0,height=7.0truecm,width=9.0truecm]{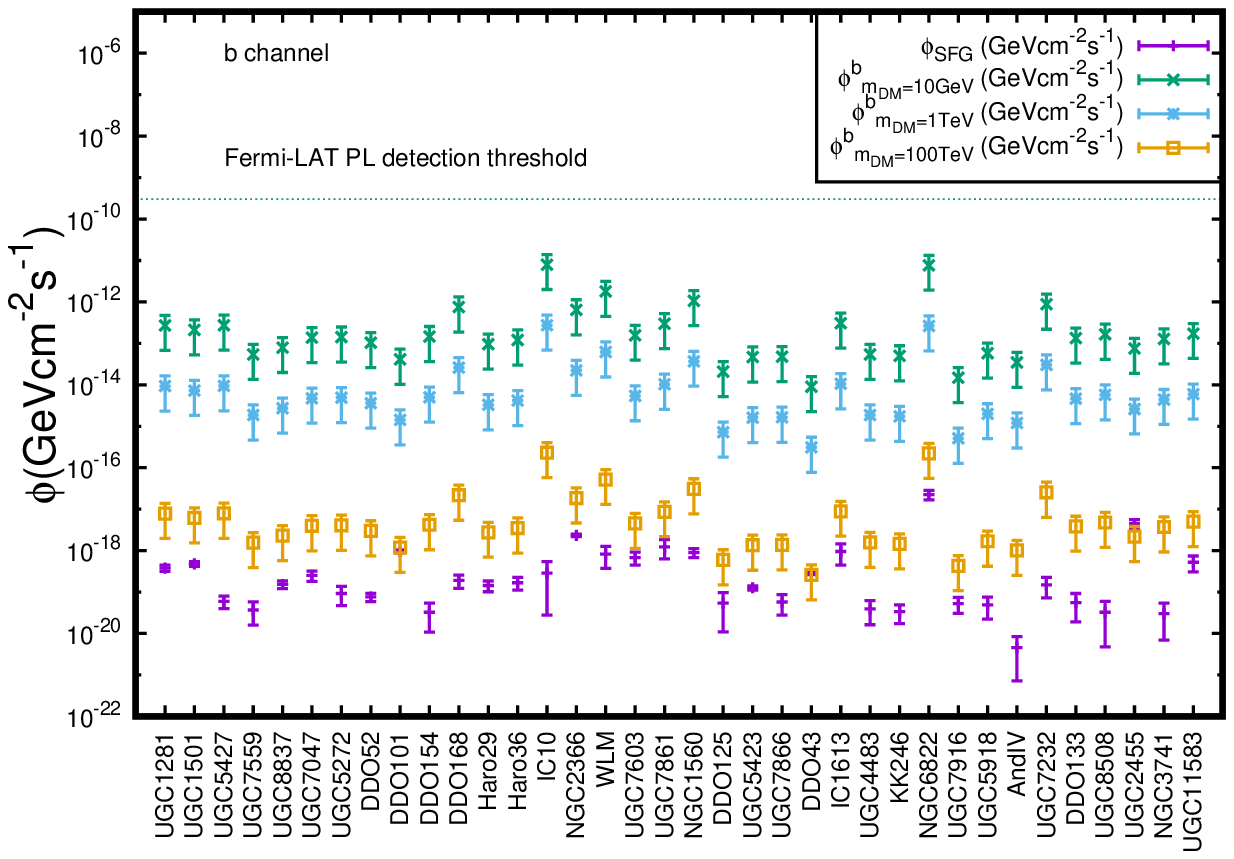}
\includegraphics[angle=0,height=7.0truecm,width=9.0truecm]{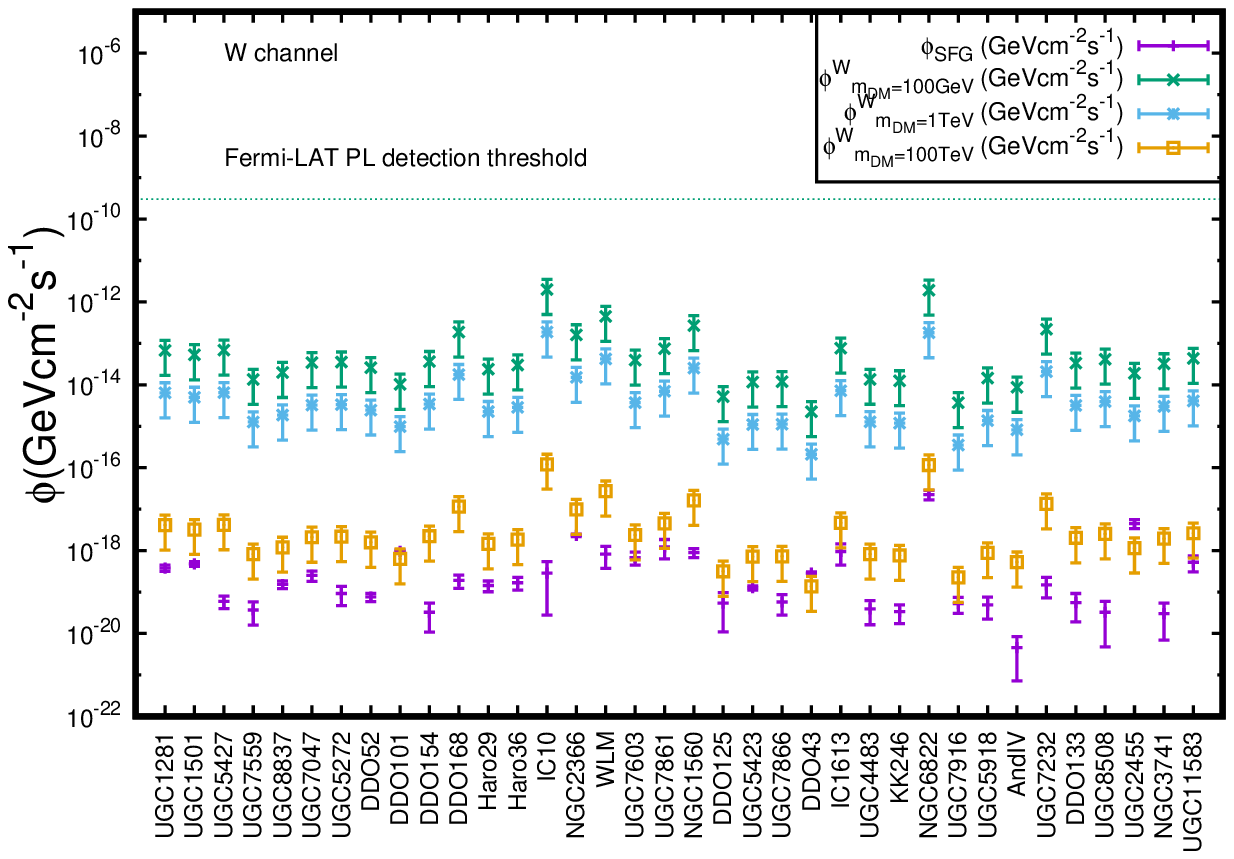}
\caption{\centerlastline \footnotesize {Gamma-ray flux $\phi^\gamma_{SFG}$ expected from the star-forming region (purple points) and DM gamma-ray fluxes $\phi^\gamma_{DM}$. From top to bottom: $\tau^+\tau^-,\, b\bar b$ and $W^+W^-$ annihilation channels and three DM masses: 10 GeV for $\tau^+\tau^-$, $b\bar b$ and 100 GeV for the $W^+W^-$ annihilation channel (green crosses); 1 TeV (blue full squares) and 100 TeV (yellow empty squares). $\hspace{10cm}$}}
\label{Lumphi}
\end{figure}

\subsection{The astrophysical gamma-ray luminosity}
\vspace{0.3cm}
In the star-forming galaxies (hereafter SFGs), the spectral energy distribution is dominated by the IR emission, that is a robust indicator of the Star Formation Rates (SFRs). Pulsars with nebula and supernovae remnants are believed to be the main accelerators of cosmic-rays that contribute to the diffuse gamma-ray emission via the interaction with gas (bremsstrahlung) or the galactic magnetic field (inverse Compton scattering). It is therefore natural to look for a correlation between the IR luminosity (then SFR) also in the star-forming regions of dIrrs, even if their luminosity is of course expected to be lower than that of starburst galaxies, but higher than that of dSphs \citep{Tosi}. In starburst galaxies the gamma-ray luminosity as function of SFR was studied by several authors, see e.g. \cite{Martin} for galaxies with SFR$\sim1$ - $10^2 \hspace{0.1cm}\mathrm{M_{\odot} yr^{-1}}$ and for SFR $\sim10^{-2} - 10^2 \hspace{0.1cm}\mathrm{M_{\odot} yr^{-1}}$ see \cite{FERMIastro}. These models are suited to better simulate normal spiral galaxies than strongly irregular ones, however it can provide insights into the effects of star formation conditions of non-thermal emissions in dIrrs. Since our dIrr galaxies have lower SFRs \citep[e.g. WLM SFR$\sim10^{-3}\hspace{0.1cm}\mathrm{M_{\odot} yr^{-1}}$, see][]{Tosi} 
their astrophysical gamma-ray emission is also expected to be lower. In fact, only bright in infrared starburst galaxies were firmly detected in gamma-rays \citep{starbust,starbust1,starbust2}. Recently, seven external SFGs have been 
detected in gamma rays by the Fermi-LAT. Six of them are rotationally supported galaxies \citep[see for references][]{fermigal,fermigal1,fermigal2,fermigal3,fermigal4,fermigal5,LMC}. Instead, fainter SFGs are most likely contributing to the IGRB \citep{Linden:2016fdd}.  
The gamma-ray luminosity $L_\gamma$ of the SFGs can be expressed as a function of their far infrared (IR) luminosity $L_\mathrm{IR}$ as:

\begin{equation}
\log_{10}\left(\frac{L_\gamma}{\mathrm{erg s}^{-1}}\right)=\alpha \log_{10}\left(\frac{L_\mathrm{IR}}{10^{10}L_\odot}\right)+\beta,
\label{Lg}
\end{equation}

\noindent
where the values of $\alpha$ and $\beta$ are $1.18\pm0.14$, $38.49\pm0.24$, respectively. These best-fit values are obtained by analyzing the gamma-ray signal that is coincident with 584 SFGs with  $L_\mathrm{IR} ~\gsim~10^8 \hspace{0.1cm} \mathrm{L_\odot}$ selected from the IRAS sample \citep[see][and references therein]{Linden:2016fdd}. 
We extrapolate this relation down to the dimmest galaxies of our sample with 
$L_\mathrm{IR}\sim\mathrm{few}\times10^6 \hspace{0.1cm} \mathrm{L_\odot}$. The values of $L_\mathrm{IR}$ for the population of 36 galaxies in our sample along with the corresponded references are listed in Table \ref{L}. We note that three dIrrs of our sample do not have the measured infrared luminosity. Therefore, in order to determine it we use galaxies in our sample with available IR luminosities and  we build a power-low relation between that and the corresponding luminosities in K-band. Where the K-band luminosities of galaxies in our sample can be found in the catalog by  \citet{karachentsev}.

\begin{table}[t!]
\centering
\begin{center}
 \resizebox{9.0cm}{!}{ 
\begin{tabular}{|c|c|c|c|c|c|}
\hline
Name & $\mathrm{L_{IR}}$& ref &$\mathrm{L_{\gamma}}$& $\mathrm{\phi^{\gamma}_\mathrm{SFG}}$ &$\mathrm{\phi^{\gamma}_\mathrm{diff}}$    \\
 & ($\mathrm{L_{\odot}}$) & $\mathrm{L_{IR}}$  &  ($\mathrm{erg\hspace{0.1cm}s^{-1}}$) & $\mathrm{(GeV\hspace{0.1cm}cm^{-2}\hspace{0.1cm}s^{-1})}$ & (cm$^{-2}\hspace{0.1cm}$s$^{-1}\hspace{0.1cm}$sr$^{-1}$)\\
\hline
UGC1281 & $6.3\times10^{7}$	&  3   & $7.8\times10^{35}$  &$3.8\times10^{-19}$ & $9.3\times 10^{-8}$ 	 \\
\hline
UGC1501  & $7.9\times10^{7}$	&   1  &$1.0\times10^{36}$ &$4.8\times10^{-19}$ & $1.0\times10^{-7}$ 	\\
\hline
UGC5427	   & $2.5\times10^{7}$ & 7   &$2.6\times10^{35}$& $6.0\times10^{-20}$	 & $5.8\times10^{-8}$ \\
\hline
UGC7559  & $7.9\times10^{6}$	&     1  &$6.8\times10^{34}$&	$3.7\times10^{-20}$ & $5.0\times10^{-8}$ \\
\hline
UGC8837	   & $5.0\times10^{7}$  &   1 &$6.0\times10^{35}$ 	& $1.5\times10^{-19}$ & $5.5\times10^{-8}$ \\
\hline
UGC7047	   & $3.2\times10^{7}$  &   1 &$3.5\times10^{35}$ 	&$2.5\times10^{-19}$ & $5.6\times10^{-8}$\\
\hline
UGC5272	     & $3.2\times10^{7}$&   1 &$3.5\times10^{35}$	 &$9.2\times10^{-20}$ & $5.9\times10^{-8}$\\
\hline
DDO52	& $5.0\times10^{7}$     &   1 &$6.0\times10^{35}$	&$7.6\times10^{-20}$ & $8.0\times10^{-8}$\\
\hline
DDO101	  & $2.5\times10^{8}$   &   1 &$4.0\times10^{36}$ 	&$9.6\times10^{-19}$& $5.5\times10^{-8}$\\
\hline
DDO154	   & $5.0\times10^{6}$  &    1&$4.0\times10^{34}$ 	&$3.2\times10^{-20}$& $4.8\times10^{-8}$\\
\hline
DDO168	   & $2.5\times10^{7}$  &   1 &$2.6\times10^{35}$ 	&$1.9\times10^{-19}$ & $5.1\times10^{-8}$\\
\hline
Haro29      & $3.2\times10^{7}$	&   1    &$3.5\times10^{35}$ 	&$1.4\times10^{-19}$ & $5.0\times10^{-8}$\\
\hline
Haro36	 & $6.3\times10^{7}$    & 1   &$7.8\times10^{35}$	&$1.7\times10^{-19}$ & $5.1\times10^{-8}$\\
\hline
IC10	       & $2.0\times10^{6}$      &   2  &$1.3\times10^{34}$ 	&$2.9\times10^{-19}$& $7.7\times10^{-7}$\\
\hline
NGC2366    & $1.3\times10^{8}$ &   1  &$1.8\times10^{36}$ 	&$2.3\times10^{-18}$& $9.0\times10^{-8}$\\
\hline
WLM	        & $6.3\times10^{6}$ 	&   1    &  $5.2\times10^{34}$ 	&$8.2\times10^{-19}$ & $7.2\times10^{-8}$\\
\hline
UGC7603 & $2.0\times10^{8}$	&    1    &$3.1\times10^{36}$	& $6.8\times10^{-19}$ & $5.5\times10^{-8}$\\
\hline
UGC7861    & $4.0\times10^{8}$ &   1  &$6.9\times10^{36}$ 	&$1.2\times10^{-18}$ & $5.7\times10^{-8}$\\
\hline
NGC1560	     & $5.0\times10^{7}$&    3  &$6.0\times10^{35}$ 	&$9.0\times10^{-19}$ & $2.1\times10^{-7}$ \\
\hline
DDO125	& $4.0\times10^{6}$	&      1       &      $3.0\times10^{34}$ &$5.4\times10^{-20}$ & $5.3\times10^{-8}$\\
\hline
UGC5423& $2.5\times10^{7}$	&      1     &      $2.6\times10^{35}$ &$1.3\times10^{-19}$& $1.1\times10^{-7}$\\
\hline
UGC7866	& $1.0\times10^{7}$	&1	&       $8.9\times10^{34}$ &$5.7\times10^{-20}$ & $5.6\times10^{-8}$\\
\hline
DDO43 & $1.6\times10^{8}$	&	4 &       $2.3\times10^{36}$ &$2.8\times10^{-19}$ & $1.0\times10^{-7}$\\
\hline
IC1613	& $4.0\times10^{6}$	& 1	&       $3.0\times10^{34}$ &$9.5\times10^{-19}$& $6.3\times10^{-8}$\\
\hline
UGC4483 & $4.0\times10^{6}$	& 1	&       $3.0\times10^{34}$ &$3.9\times10^{-20}$ & $7.7\times10^{-8}$\\
\hline
KK246	& $1.3\times10^{7}$	& 7	&       $1.2\times10^{35}$ &$3.3\times10^{-20}$& $2.8\times10^{-7}$\\
\hline
NGC6822	 & $4.0\times10^{7}$	& 5	&       $4.6\times10^{35}$ &$2.3\times10^{-17}$& $3.4\times10^{-7}$\\
\hline
UGC7916	& $2.5\times10^{7}$	&	1 &       $2.6\times10^{35}$ &$5.3\times10^{-20}$& $5.3\times10^{-8}$\\
\hline
UGC5918	 & $2.0\times10^{7}$	& 1	&       $2.0\times10^{35}$ &$4.9\times10^{-20}$& $5.1\times10^{-8}$\\
\hline
AndIV	& $2.5\times10^{6}$	& 7	&       $1.8\times10^{34}$ &$4.5\times10^{-21}$& $1.1\times10^{-7}$\\
\hline
UGC7232	 & $1.0\times10^{7}$	& 1	& 	 $8.9\times10^{34}$ &$1.5\times10^{-19}$& $6.1\times10^{-8}$\\
\hline
DDO133	& $1.6\times10^{7}$	& 1	&       $1.5\times10^{35}$ &$5.5\times10^{-20}$ & $5.2\times10^{-8}$\\
\hline
UGC8508 & $2.5\times10^{6}$	& 1	&        $1.8\times10^{34}$ &$3.2\times10^{-20}$& $5.4\times10^{-8}$ \\
\hline
UGC2455 & $1.0\times10^{9}$	& 5	&       $2.0\times10^{37}$ &$4.5\times10^{-18}$& $2.2\times10^{-7}$\\
\hline
NGC3741 & $3.2\times10^{6}$	& 1	&       $2.3\times10^{34}$ &$3.0\times10^{-20}$& $6.5\times10^{-8}$\\
\hline
UGC11583 & $1.0\times10^{8}$& 6	&     $1.4\times10^{36}$  &$5.2\times10^{-19}$& $3.1\times10^{-7}$\\
\hline
\hline
\end{tabular}
}
\end{center}

\caption{\centerlastline \footnotesize{Infrared luminosities $\mathrm{L_{IR}}$, $\mathrm{L_{IR}}$ reference, gamma-ray luminosity $\mathrm{L_\gamma}$, gamma-ray fluxes $\mathrm{\phi^{\gamma}_\mathrm{SFG}}$ and diffuse background flux contribution $\mathrm{\phi^{\gamma}_\mathrm{diff}}$. $\mathrm{L_{IR}}$ references: 1~-~\citet{dale09}, 2~-~\citet{brauher08}, 3~-~\citet{IRAS92}, 4~-~\citet{lisenfeld06}, 5~-~\citet{sanders03}, 6~-~\citet{ABRAHAMYAN15}, 7~-~predicted  using the luminosity in K-band (see text for the details).  $\hspace{10cm}$ }} 
\label{L}
\end{table}

\begin{table}[h!]
\centering
\begin{center}
 \resizebox{9.0cm}{!}{ 
\begin{tabular}{|c|c|c|c|c|c|}
\hline
$\mathrm{m}_\mathrm{DM}$   &   10 GeV & 100 GeV & 1 TeV & 10 TeV & 100 TeV   	\\
\hline
\hline
$\mathrm{P}^{\tau^+\tau^-}$ 	&$5.8\times10^{-29}$ 	&$7.4\times10^{-30}$ &$1.5\times10^{-31}$ 	 & $1.1\times10^{-33}$	& $2.1\times10^{-35}$ 			\\
\hline
$\mathrm{P}^{b\bar b}$ 	&$3.2\times10^{-29}$ 	&$1.1\times10^{-29}$ &$1.1\times10^{-30}$ 	 & $4.0\times10^{-32}$	& $9.2\times10^{-34}$ 		\\
\hline
$\mathrm{P}^{W^+W^-}$ & $-$ 	&$8.0\times10^{-30}$ &$7.5\times10^{-31}$ 	 & $2.4\times10^{-32}$	& $4.9\times10^{-34}$ 				\\

\hline
\hline
\end{tabular}
}
\end{center}
\caption{\footnotesize{P-part ($\mathrm{GeV}^{-1}\mathrm{cm}^3\mathrm{s}^{-1}$) for annihilation events in a thermal WIMP scenario ($<\sigma v>=3\times10^{-26}\mathrm{cm}^{-3}\mathrm{s}^{-1}$).
 $\hspace{2.7cm}$}}

\label{Pfactor}
\end{table}

In Table  \ref{L} we also provide the values of $L_\gamma$  calculated by using Eq. \ref{Lg} and the gamma-ray fluxes of each dIrr galaxy calculated as: 

\begin{equation}
\phi_\mathrm{SFG}^\gamma=\frac{L_\gamma}{4\pi d^2}\,\,,
\label{gammaflux}
\end{equation}
\noindent
where $d$ is the distance to the target, given in Table \ref{Tab:Irr}. \\

In Fig. \ref{Lumphi} we show the astrophysical gamma-ray emission and the Fermi-LAT detection threshold. The gamma-ray fluxes $\phi^\gamma_{SFG}$ associated with the star-forming regions of dIrrs in our sample are smaller than $10^{-17}\mathrm{ph}\,\mathrm{cm}^{-2}\,\mathrm{s}^{-1}$ resulting largely compatible with the zero background hypothesis. Their uncertainties are obtained as a combination of both the uncertainties on the distance of the target and of the $\alpha$, $\beta$ parameters in Eq. \ref{Lg}. 

\subsection{The dark matter gamma-ray luminosity}
\vspace{0.3cm}

In this section we will focus on the particle physics P-part of Eq. \ref{flux}. Within the hypothesis of a thermal WIMP candidate ($<\sigma v>=3\times10^{-26}\mathrm{cm}^{3}\mathrm{s}^{-1}$), the P-part only depends on the DM particle mass and the annihilation channel. Therefore, we calculate the P-part in Eq. \ref{Eq:P} for three different annihilation channels ($\tau^+\tau^-, b\bar b$ and $W^+W^-$) and five values of DM mass ($10,100,10^3,10^4,10^5$ GeV) integrated on the energy range of interest (1-100 GeV). This energy range is in agreement with the estimated astrophysical gamma-ray flux $\phi^\gamma_{SFG}$ of the star-forming region, as discussed above. The P-part is given in Table \ref{Pfactor}. We adopt the J-factors for the Fermi-LAT PSF, that is $\theta\approx0.5^\circ$, given in Table \ref{angles}. For each dIrrs, the maximum (minimum) value of the DM-related integrated flux is given by the $\tau^+\tau^-$ annihilation channel and the $10$ GeV ($100$ TeV) DM particle mass. On Fig. \ref{Lumphi} we compare the expected gamma-ray fluxes that originate from the star-forming regions of dIrrs with that expected from the annihilation of thermal WIMP particles for a given channel and a DM particle mass. 
The uncertainty on $\phi^\gamma_{DM}$ includes the $75\%$ error on the J-factor. The Fermi-LAT detection threshold in Fig. \ref{Lumphi} is given by \cite{Linden:2016fdd}. We can then conclude: (i) the estimated emission of the DM-related gamma-ray flux $\phi^\gamma_{DM}$ enclosed in Fermi-LAT PSF can be up to 8 orders of magnitude higher, depending on the DM mass, than the astrophysical gamma-ray flux $\phi^\gamma_{SFG}$ associated with the star-forming region of the same dIrr. Therefore, the \textit{extra} astrophysical gamma-ray background is negligible in these objects, and we can apply to dIrrs the same analysis used to study \textit{classical} dSphs galaxies; (ii) the secondary $\phi^\gamma_{DM}$ emitted by the annihilation events of the thermal DM particles in dIrrs remain below the Fermi-LAT detector sensitivity threshold. In the following section we will set upper limits on the DM particle mass and annihilation cross-section via the first analysis of dIrr galaxies. \\

We conclude that the astrophysical gamma-ray emission in dIrr galaxies can be considered negligible similar to what is assumed in dSphs. The latter statement is valid for, at least, a DM particle mass of $10-10^4$ GeV and the energy range studied in this work. For this limited energy range, the expected signal from a DM particle candidate of 10-100 TeV would in fact be contaminated by the astrophysical gamma-ray contribution. 

\section{Sensitivity analysis}
\label{analysissecVII}

In the previous sections we have theoretically demonstrated that dIrrs could be considered as objects with approximately zero astrophysical gamma-ray background and with the DM halo that appears as a point-like or a marginally extended source for gamma-ray telescopes. As a first application of such a theoretical rationale, we are going to place constraints on the DM particle mass and annihilation cross-section  with our sample of galaxies and by means of the sensitivity study of the Fermi-LAT detector. We ask for detection of secondary gamma rays produced in the DM halo of dIrr galaxies with statistical significance of $5\sigma, 3\sigma$ and $1\sigma$:
\begin{equation}
\chi=\frac{\phi_\gamma^\mathrm{DM}\sqrt{\Delta\Omega_\mathrm{PSF} A_\mathrm{eff} t_\mathrm{exp}}}{\sqrt{\phi_\gamma^\mathrm{DM}+\phi_\mathrm{Bg}}}> 5,\,3,\,1 .
\label{X}
\end{equation}

The effective area $A_\mathrm{eff}$ and exposition time $t_\mathrm{exp}$ for the Fermi-LAT detector used in this analysis are given in Table \ref{telescopes}. We fix the exposition time $t_\text{exp}$ and, as first approximation, we keep constant the effective area $A_\text{eff}$, although the latter also depends on the energy range, the incidence angle and the azimuthal angle \cite{Pv6}. These dependences would affect a proper data analysis with respect to these first results. Notice that, the solid angle $\Delta\Omega_\mathrm{PSF}=2.4\times10^{-4}\mathrm{sr}$ corresponding to the Fermi-LAT PSF of $0.5^\circ$, allows to include the biggest parts of galaxies' DM halos in more than 70\% of galaxies in the sample. In fact, the J-factors calculated on the Fermi-LAT PSF are identical to those calculated on the whole virial angular dimensions of the associated DM haloes. However, we will show that assuming $\Delta\Omega_\mathrm{PSF}=10^{-5}\mathrm{sr}$ (PSF of $0.1^\circ$) we get an improvement in the results by almost a factor of 8, depending on the particular source (see next subsection and Appendix \ref{AppC} for details). \\

Showing that the astrophysical gamma-ray contamination in dIrr is negligible, we estimate the gamma-ray background $\phi_\mathrm{Bg}$ in Eq. \ref{X} taking into account both the galactic interstellar diffuse emission model designed to be used for point source analysis ($gll\_iem\_v06.fits$) and the isotropic spectral template ($ISO\_P8R2\_SOURCE\_V6\_v06.txt$). The latter includes both extragalactic diffuse gamma rays and the remaining residual (misclassified) cosmic-ray emission from a fit to the all-sky emission with |b|>30 deg, that are not represented in the Galactic diffuse model. In particular the SOURCE class provides good sensitivity for analysing the point sources and moderately extended sources. Therefore, for each k-dIrr galaxy in the sample we assume:

\begin{equation}
\phi^k_\mathrm{Bg}=\phi^k_\mathrm{diff}+\phi_\mathrm{iso},
\end{equation}

\noindent
although the diffuse emission above $10$ GeV is considered to be sub-dominant with respect to the isotropic contribution \citep{PhysRevLett.103.251101}. We then calculate the diffuse emission background contribution between $1-100$ GeV for the position of each dIrr in the sky. These values are given in Table \ref{L} and the integrated isotropic contribution is estimated to be $\phi_\mathrm{iso}=5.0\times10^{-7}\mathrm{cm}^{-2}\mathrm{s}^{-1}\mathrm{sr}^{-1}$ between $1-100$ GeV. We also check the Fermi-LAT 3FGL Catalog \citep{Acero:2015hja} and we exclude Haro36 dIrr that is spatially coincident with the 3FGLJ1248.0+5130 BL LAC object within $0.5^\circ$ degrees in the Fermi-LAT sky. In what follows we show the results of a single galaxy study as well as of the stacked analysis. 

\vspace{0.5cm}

\subsection{Individual limits}

%\vspace{0.3cm}

\begin{figure}[t]
\begin{center}
{\includegraphics[angle=0,height=7truecm,width=9truecm]{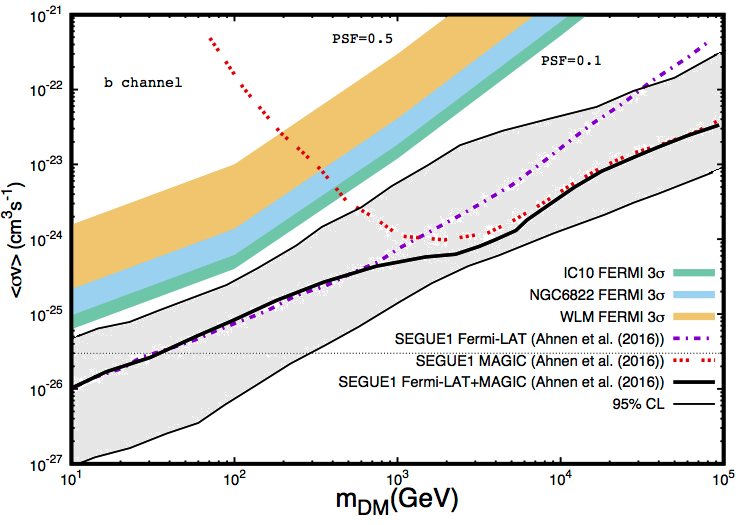}}
\caption {\centerlastline \footnotesize {Upper limits for the DM particle mass and annihilation cross section for the $b\bar b$ annihilation channel and the three best candidates of our sample, that are NGC6822 (blue), IC10 (green) and WLM (orange). The shadowed areas show how the constraints improve if we change a PSF from $0.5^\circ$ to $0.1^\circ$, keeping the $3\sigma$ statistical significance. The colour regions also include the $43\%$ of uncertainty related to the error on the J-factors. Additionally we  show the upper limits for the SEGUE1 dSph obtained by the combined analysis of Fermi-LAT and MAGIC data (solid black line) within the $95\%$ of CL (gray region) \citep{Ahnen:2016qkx}. \hspace{8cm}
}
}
\label{Bests}
\end{center}
\end{figure}

In Fig. \ref{Bests} we show the best results obtained within our sample of galaxies, that are given by the study of NGC6822, IC10 and WLM dIrrs. Notice that NGC6822 galaxy group is excluded in \cite{Lisanti:2017qoz} because of its location that is less than $2^\circ$ from the center of another galaxy group. Here, instead, we keep it since we are only considering the central galaxy which does not overlap with any source in the $0.1/0.5$ degrees of interest in our analysis. Whereas the Fermi-LAT PSF is strongly energy dependent, we show, as shadowed regions, the improvement on the upper limits when we switch the point-like analysis from an angular resolution of $0.5^\circ$ to $0.1^\circ$ with the statistical significance of $3\sigma$, in both cases. This can help to graphically visualize the possible inaccuracy that could affect the sensitivity study with respect to a proper data analysis. 
  
In the same Fig. \ref{Bests} we also show the upper limit within $95\%$ of confidence level (hereafter denoted as CL) (gray region) obtained for the SEGUE1 dSph galaxy with the combined analysis of 6-years of Fermi-LAT and 158 hours of MAGIC observations \citep{Ahnen:2016qkx}. Let us notice that the most stringent upper limits we get are in the case of IC10 dIrr galaxy. Although, these limits are still weaker, by a factor of a few, than that of SEGUE1 dSph galaxy, the assumed value for the SEGUE1 J-factor ($\sim 10^{19.5\pm0.29}\,\mathrm{GeV}^2\mathrm{cm}^{-5}$ on $0.5^\circ$) is largely higher than the lower limit ($\sim10^{15}\,\mathrm{GeV}^2\mathrm{cm}^{-5}$) when accounting for the systematical uncertainties on its triaxiality \citep{triaxSph}, as shown in Fig.~\ref{thetaEXPJ}-\ref{thetavirJ}. In other words, the SEGUE1 constraints based on the lower bound of the J-factor value would be comparable with the limits of this work. On the contrary, the 3$\sigma$ uncertainties of the dIrrs' upper limits do include $43\%$ of uncertainties related with the corresponding uncertainties on the J-factors (see Eq. (\ref{X})). In Appendix \ref{AppC} we show the upper limits on the DM particle mass and annihilation cross-section for each galaxy in our sample.

\subsection{Stacked limits}

In this Section we illustrate the predictions for the upper limits on the DM annihilation cross-section from the combined/stacked analysis of sub-sample of galaxies. We notice that the most promising combination is given by a sub-sample of 7 targets, that are: DD0168, IC10, NGC2366, WLM, NGC1560, NGC6822, UGC7232. Then we define:

\begin{equation}
\phi_\gamma^\mathrm{stack-DM}=\sum_{k=1}^{7}<J>_{\Delta\Omega}^k {\Delta\Omega^\mathrm{PSF}} P
\end{equation}

\noindent
where $P$ is the particle physics factor defined in Eq. (\ref{Eq:P}). The backgrounds are defined as:

\begin{equation}
\phi_\mathrm{Bg}^\mathrm{stack}=\sum_{k=1}^{7}\phi^k_\mathrm{Bg}.
\end{equation}
 
\begin{figure}
\centering
\includegraphics[angle=0,height=6.5truecm,width=9.0truecm]{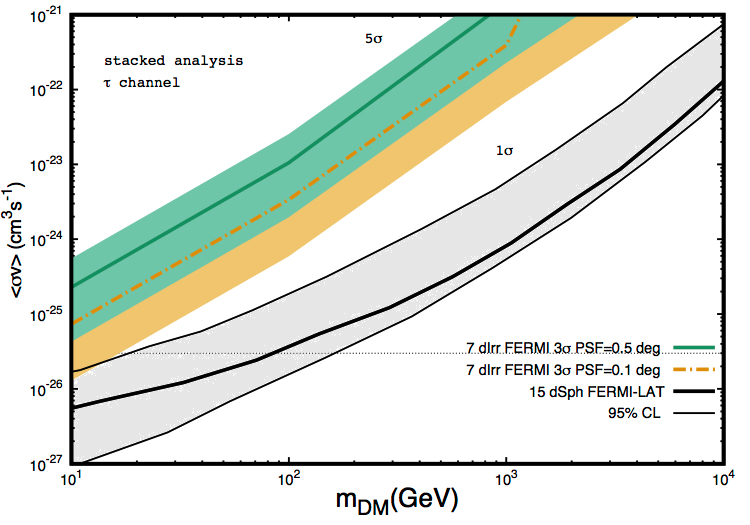}
\includegraphics[angle=0,height=6.5truecm,width=9.0truecm]{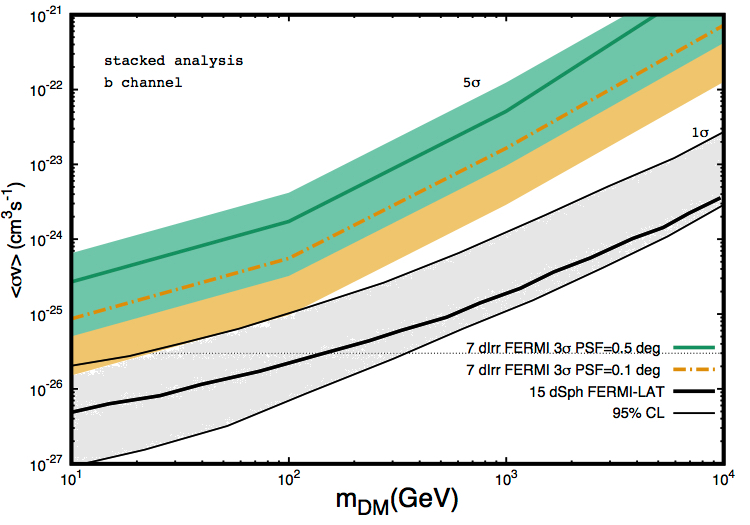}
\includegraphics[angle=0,height=6.5truecm,width=9.0truecm]{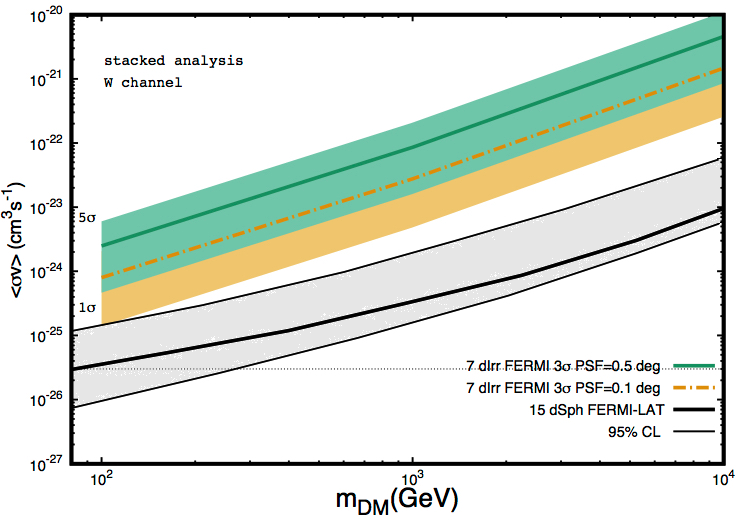}
\caption{\centerlastline \footnotesize Sensitivity study of the DM particle mass and annihilation cross-section by the combined analysis of 7 dIrr galaxies and a PSF of both $0.5^\circ$ (green band ($5\sigma-1\sigma$ region) and solid line ($3\sigma$)) and $0.1^\circ$ (orange band ($5\sigma-1\sigma$ region) and dot-dashed line ($3\sigma$)). The black solid line shows the $2\sigma$ upper limits within $95\%$ of CL (gray region)  for the combined analysis of 15 dSph galaxies that was performed by the Fermi-LAT collaboration \citep{dSphFERMI, dSphFERMI1}.\hspace{3cm}}
\label{stacked}
\end{figure}

The resulting exclusion limits are shown in Fig. \ref{stacked}. In order to take into account the PSF variability with the energy, we show the results of the point-like analysis for both a PSF of $0.5^\circ$ (green band ($5\sigma-1\sigma$ region) and line ($3\sigma$)) and $0.1^\circ$ (orange band ($5\sigma-1\sigma$ region) and line ($3\sigma$)). In the same figure, we also show the $2\sigma$ significance limits within the $95\%$ of CL  obtained by the analysis of 15 dSph galaxies by the Fermi-LAT collaboration \citep{dSphFERMI1}. We note that although our upper limits are not necessary better than that obtained by the analysis of dSph galaxies, they could be improved by the discovery of promising dIrr targets.
 
 \vspace{0.5cm}
\section{Discussion}
\label{discussion}

\begin{figure}
\centering
\includegraphics[angle=0,height=6.5truecm,width=9truecm]{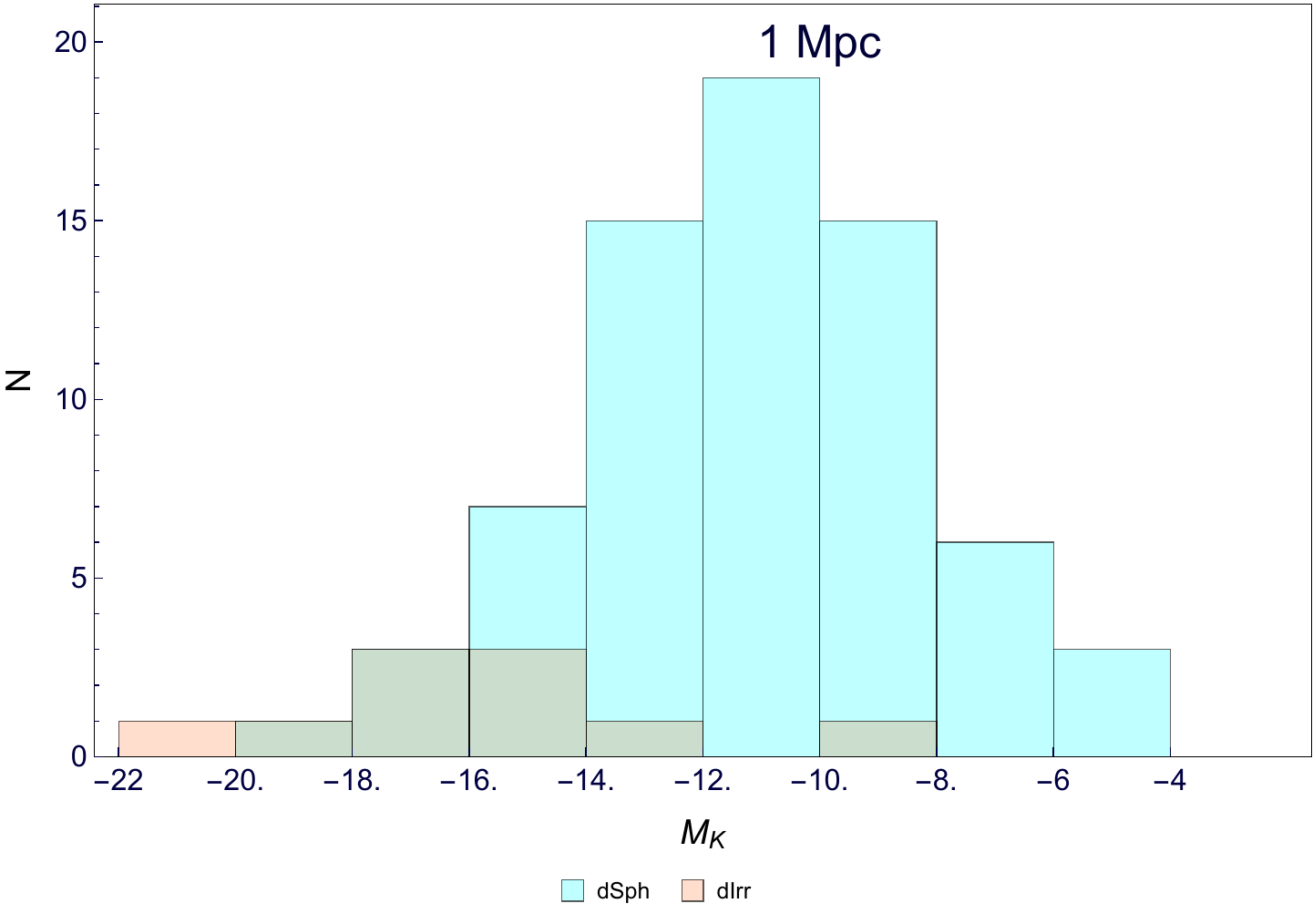}
\includegraphics[angle=0,height=6.5truecm,width=9truecm]{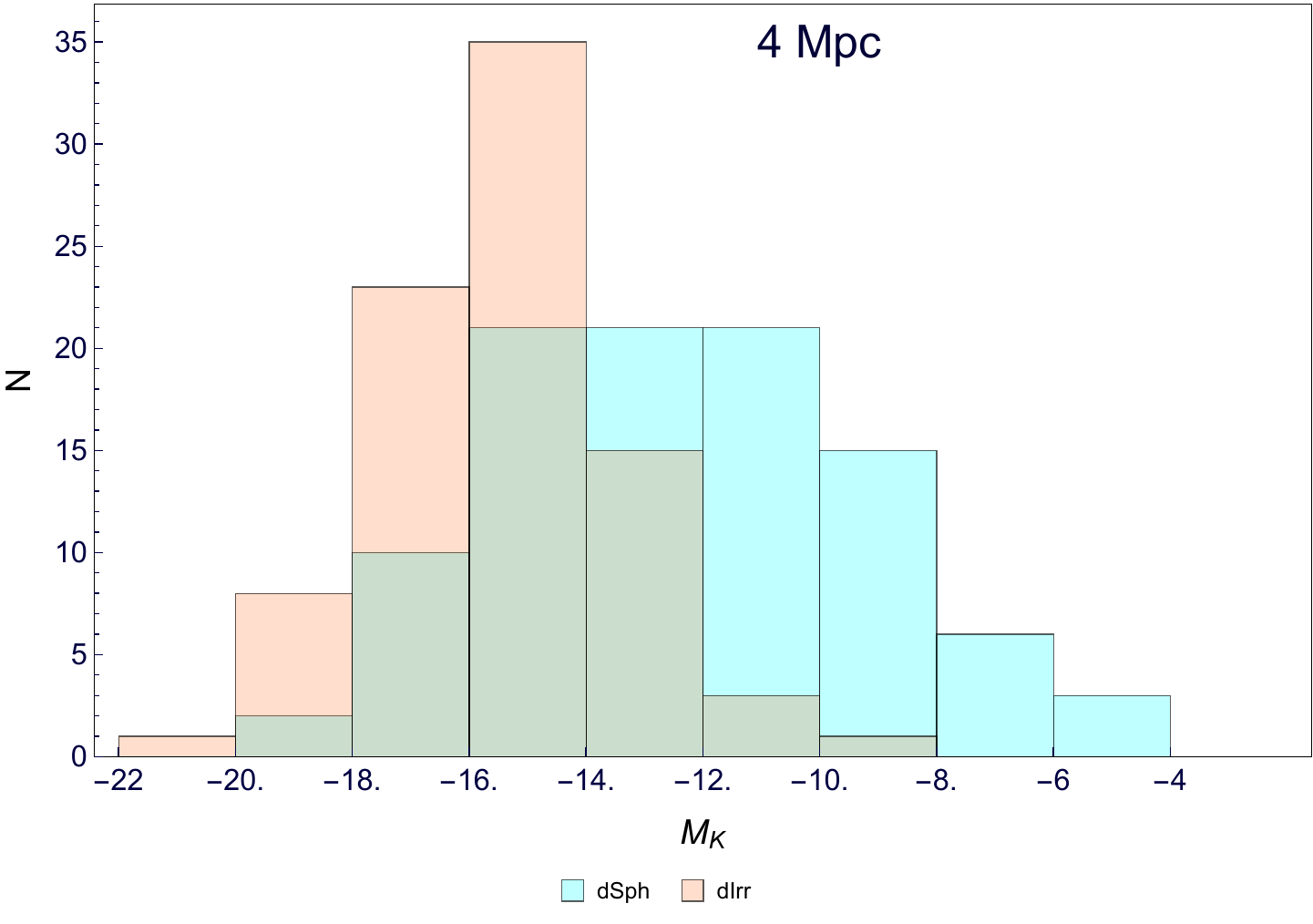}
\includegraphics[angle=0,height=6.5truecm,width=9truecm]{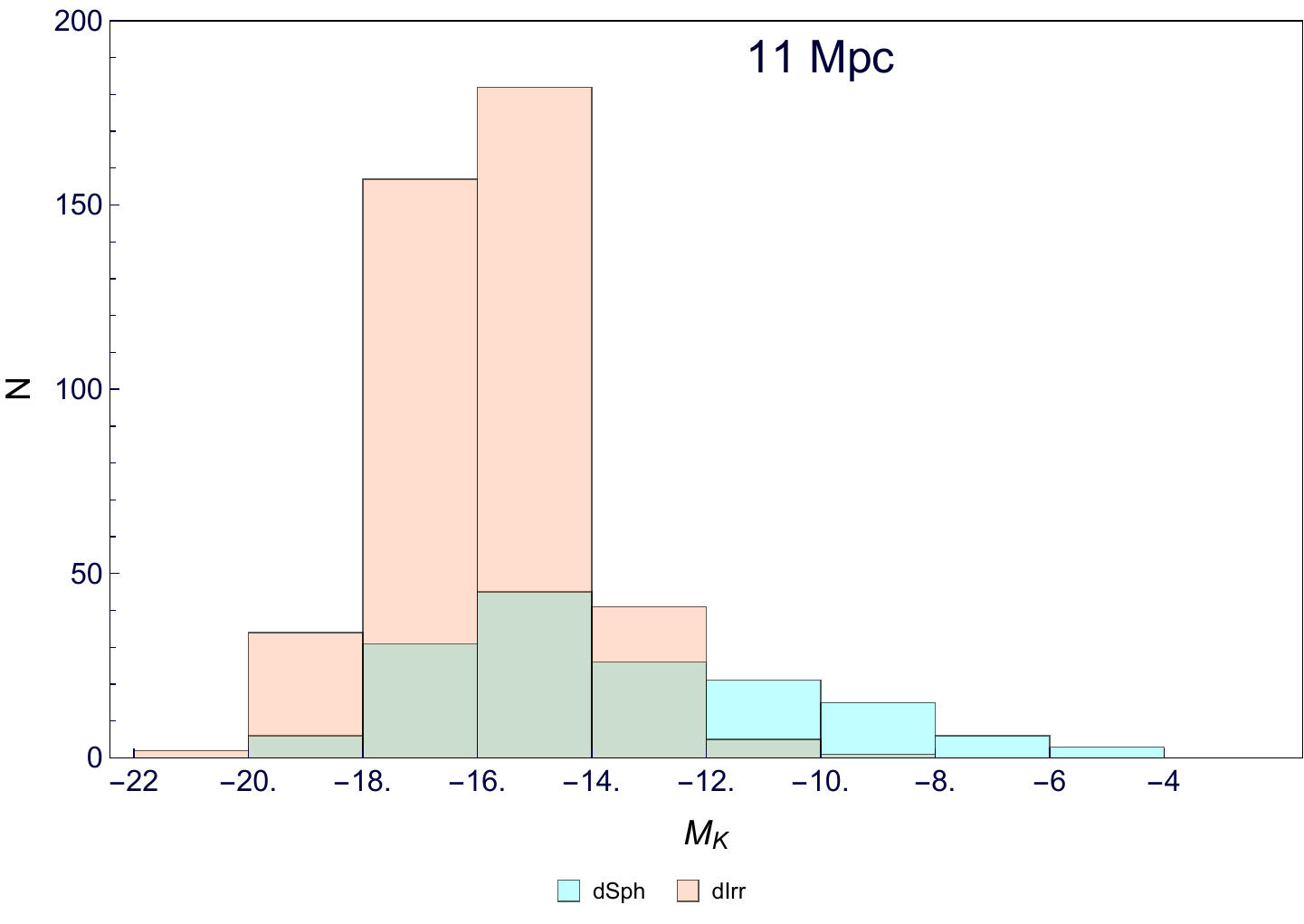}
%\vspace{2.cm}
\caption{\centerlastline \footnotesize {Distribution of dSphs (blue) and dIrrs (pink) according to their absolute K-band magnitude for 3 distance bins: 1 Mpc (upper panel), 4 Mpc (central panel), 11 Mpc (bottom panel). The green color shows overlapping regions \citep[data are taken from][]{karachentsev}. \hspace{5cm} }}
\label{N}
\end{figure}

The purpose of this paper is to analyze dIrr galaxies as new targets for indirect DM searches and %to give
estimate the constraints that can be set on the DM particle mass and annihilation cross-section by new prospective experimental data analysis for such class of objects. Although our results show that dIrr are of one order of magnitude less competitive with respect to dSph galaxies, we would like to stress that the number of detected dIrr galaxies in the Local Volume is greatly increasing with time \citep[see][]{karachentsev}. In Fig. \ref{N} we plot histograms that show the distribution of dSphs and dIrrs according to their absolute K-band magnitude for 3 distance bins \citep[data are taken from][]{karachentsev}. The well-studied Milky Way dSph galaxies are enclosed in a radius of $\lsim 300$ kpc \cite{Fermi-LAT:2016uux}, and within $d < 1$ Mpc the number of dSphs is almost $20\%$ bigger than that of dIrrs (upper panel in Fig. \ref{N}). Starting from the distance of $\sim4$~Mpc (middle panel in Fig\ref{N}) dIrrs start to outnumber dSphs. The 7 dIrrs included in the selected sub-sample for the combined study reach a distance of $\sim5$ Mpc, but the sample only includes galaxies of a particular region of the sky (see the sky map in Fig. \ref{skymap}). 
Therefore, the increasing number of dIrr detected and studied in their kinematics may result in placing competitive bounds with respect to those derived from the stacked analysis of the limited number of dSph galaxies. In fact, the total number of dIrrs in the Local Volume is expected to be $\sim200\%$ of that of dSphs (bottom panel in Fig.\ref{N}). Moreover, the kinematics of closer and fainter rotationally-supported galaxies, such as Low Surface Brightness (LSB) galaxies have been recently studied \cite{LSB} and they are also expected to occupy massive DM halos.

\section{Conclusions}
\label{conclusions}
In this paper, we analyze a sample of 36 dIrr galaxies as new point-like targets for indirect DM searches with gamma-ray telescopes. These DM dominated rotationally supported galaxies, differently from pressure supported dSphs, allow us to well constrain their DM density distribution profiles by means of the accurate kinematics and the universality in the DM distribution. The kinematics of these galaxies points to the existence of the DM core profiles with DM halos extending far beyond their optical star-forming regions. Although a gamma-ray background contribution of astrophysical origins is expected in these regions, it is assumed to be localized in a very inner region that stays unresolved by the current generation of gamma-ray telescopes. First of all, we calculate the astrophysical factors for a sample of 36 dIrr galaxies in the Local Volume ($<11$ Mpc) and we show that the uncertainties on the J-factors, although estimated to be bigger than $75\%$, are, in most cases, better than that of the \textit{classical} dSph galaxies. Two galaxies in the sample (NGC6822 and IC10) have J-factors $\gsim \mathrm{few}\times10^{17}\hspace{0.1cm}\mathrm{GeV}^2\mathrm{cm}^{-5}$ and five more (DD0168, NGC2366, WLM, NGC1560 and NGC7232) bigger than $ \mathrm{few}\times10^{16}\hspace{0.1cm}\mathrm{GeV}^2\mathrm{cm}^{-5}$. Secondly, we show that the spatially localized astrophysical background contribution is expected to remain unresolved, therefore negligible with respect to the diffuse and isotropic gamma-ray background in the 1-100 GeV energy range. On the other hand, the DM halo in dIrrs could appear as point-like or  marginally extended for gamma-ray telescopes (Fig. \ref{angularres}). We study the possibility to set competitive constraints on the DM particle mass and annihilation cross-section via the study of this category of galaxies. 
The upper limits for each individual galaxy are less stringent than that of the analysis of the SEGUE1 dSph performed by the Fermi-LAT and MAGIC collaborations. However, notice that the last statement is true within a certain assumption on the value of the SEGUE1 J-factor, where its lower J-factor bound can be up to 4 orders of magnitude smaller than the commonly assumed value. Additionally, let us notice that the claim for a $2\sigma$ detection of the DM signal in ultra-faint dwarf Indus II ($\mathrm{m}_\mathrm{DM}\approx 100$ GeV, $<\sigma v>\approx10^{-23}$, 6 years, $2.4\sigma$) \citep{Fermi-LAT:2016uux} should be confirmed by the analysis of at least three galaxies in our sample, that are NGC6822, IC10, and the WLM galaxy. On the other hand, the claim for the DM detection in Tucana~III ($\mathrm{m}_\mathrm{DM}\approx 100$ GeV, $<\sigma v>\approx10^{-26}$, 7 years, $2.4\sigma$) \citep{ShangLi} cannot be confirmed, following our results.
We also show the upper limits for the stacked analysis of a selection of 7 dIrr galaxies and a PSF of $0.5^\circ$ and $0.1^\circ$, that
 starts to approach the constraints given by the analysis of 15 dSphs with Fermi-LAT. 
Considering that this analysis makes use only of the Fermi-LAT background models, the comparison with the constraints obtained through the proper data analysis of the \textit{classical} dSphs is not straightforward \citep{dSphFERMI,dSphFERMI1} and further investigation is needed.\\

Altogether, in this paper, by means of a theoretical approach, we reconsider dIrr galaxies as new point-like targets for DM searches. This proof of concept could open new avenues concerning the data analysis of promising dIrrs. These galaxies could potentially represent complementary targets for DM searches because of their different kinematics, morphology and cosmological evolution, that result in density profiles that are different from that of dSphs. We would like to mention that the next generation large-area HI surveys, including SKA, will help to increase the number of the detected late-type galaxies in the Local Universe, as well as it will improve the kinematical measurements of their HI content (see e.g. \citep{2015arXiv150603474A}). Therefore, the increasing number of dIrr galaxies with measured kinematics could improve the significance of the results.
Moreover, further investigation will be addressed to the study of this class of galaxies at higher energy scale. In fact, because AGN are not observed in dIrr galaxies, the astrophysical gamma-ray background is expected to be negligible even at very high energy. Preliminary results for these objects with two-years data of the HAWC observatory have been already obtained for the TeV energy scale \citep[see ][ for more details]{Cadena:2017ldx}.\\
\\

\vspace{-1cm}
\begin{acknowledgments}
This work has been supported by QGSKY, by the Agencia Estatal de Investigaci\'on (AEI) y al Fondo Europeo de Desarrollo Regional (FEDER) FIS2016-78859-P(AEI/FEDER, UE), by the MINECO (Spain) project FIS2014-52837-P and Consolider-Ingenio MULTIDARK CSD2009-00064, and partially by the H2020 CSA Twinning project No.692194 ORBI-T-WINNINGO. The authors are grateful to
the anonymous referee for useful comments and suggestions. VG thanks the support of the Spanish Red Consolider MultiDark FPA2017-90566-REDC and she is also grateful to P. Ullio, A. Urbano, M. A. Sanchez-Conde, N. Fornengo, G. Gomez-Vargas, G. Zaharijas, R. Alfaro and S. Hernandez, for useful discussions, and to V. Avila-Reese, A.G.X. Gonzalez-Morales and O. Valenzuela for interesting considerations related with the Appendix \textit{Details on the J-factor}.  
EK work was supported by the S\~ao Paulo Research Foundation (FAPESP) under the grant \#2016/26288~-~9.

\end{acknowledgments}

\bibliography{biblio}

\appendix

\appendix{
\renewcommand\thefigure{\thesection.\arabic{figure}}    
\section{Rotation curves}
\label{AppA}

As it was mentioned in Section \ref{Irr} the URC in most of the cases is able to fit the individual rotation curves of galaxies in our sample. On Fig.~\ref{v12} we plot the \textit{prediction} for the DM density distribution profile for each galaxy (as obtained by the URC studies) over kinematical measurements of each. The difficulty for three galaxies of the sample (NGC6822, AndIV and UGC8508) to reproduce the outer rotation curves may be due to the uncertainties on the estimated inclinations or disk length scales of the objects.}

\begin{figure*}[h!]
\begin{center}
\includegraphics[angle=0,height=18.8truecm,width=12.8truecm]{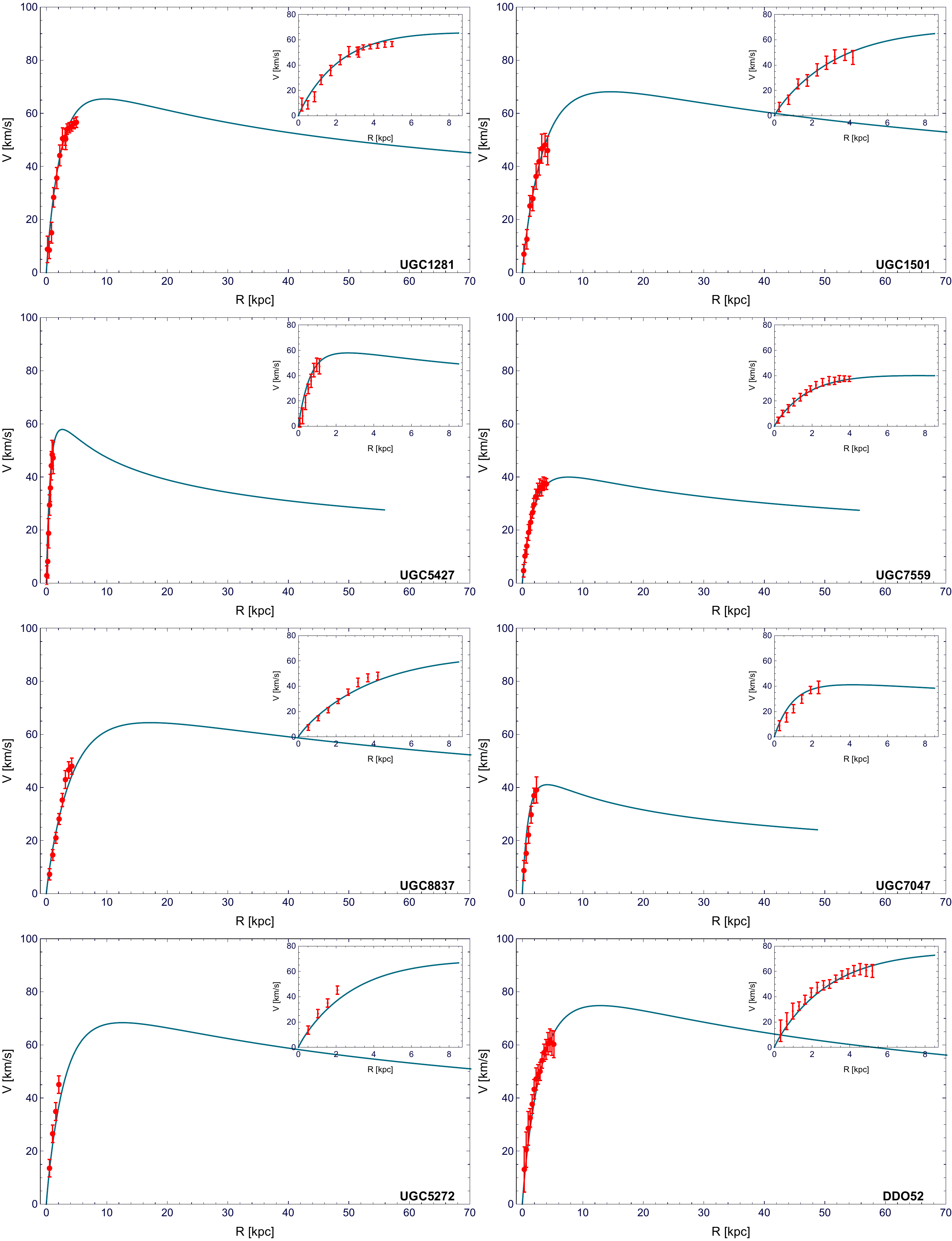}

\caption {\centerlastline \footnotesize{Red points with error bars represent the observed circular velocities of galaxies in the sample. Blue lines represent the \textit{predictions} for the circular velocities from the universal rotation  curve analysis (plotted up to the virial radii of galaxies). We also show the zoomed-in area for the inner 8 kpc.  $\hspace{1.2cm}$}}
\label{v12}
\end{center}
\end{figure*}

\newpage

\begin{figure*}[h!]
\begin{center}
\includegraphics[angle=0,height=18.8truecm,width=12.8truecm]{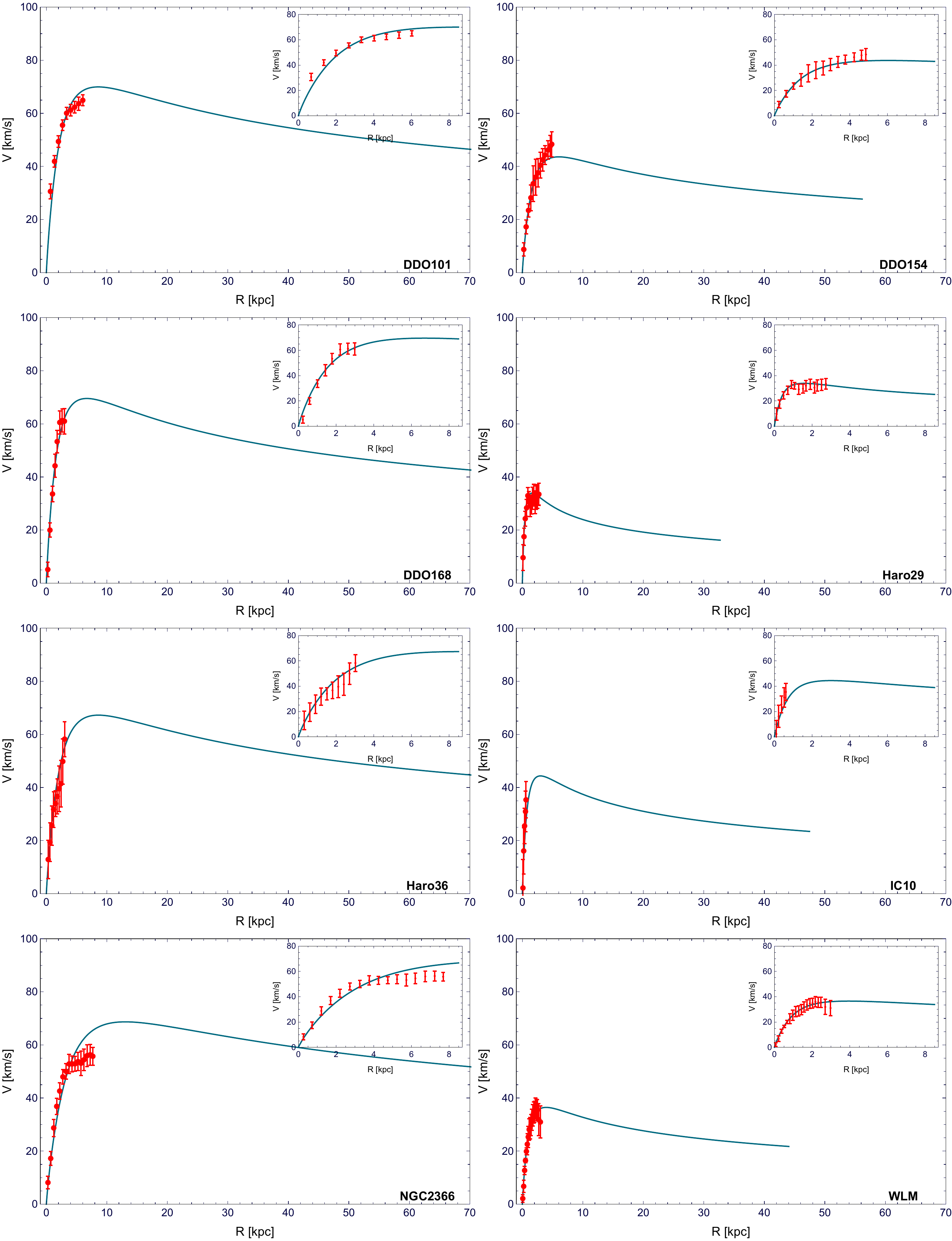}
\contcaption{\footnotesize{\it{continued}}}
\end{center}
\end{figure*}

\newpage

\begin{figure*}[h!]
\begin{center}
\includegraphics[angle=0,height=18.8truecm,width=12.8truecm]{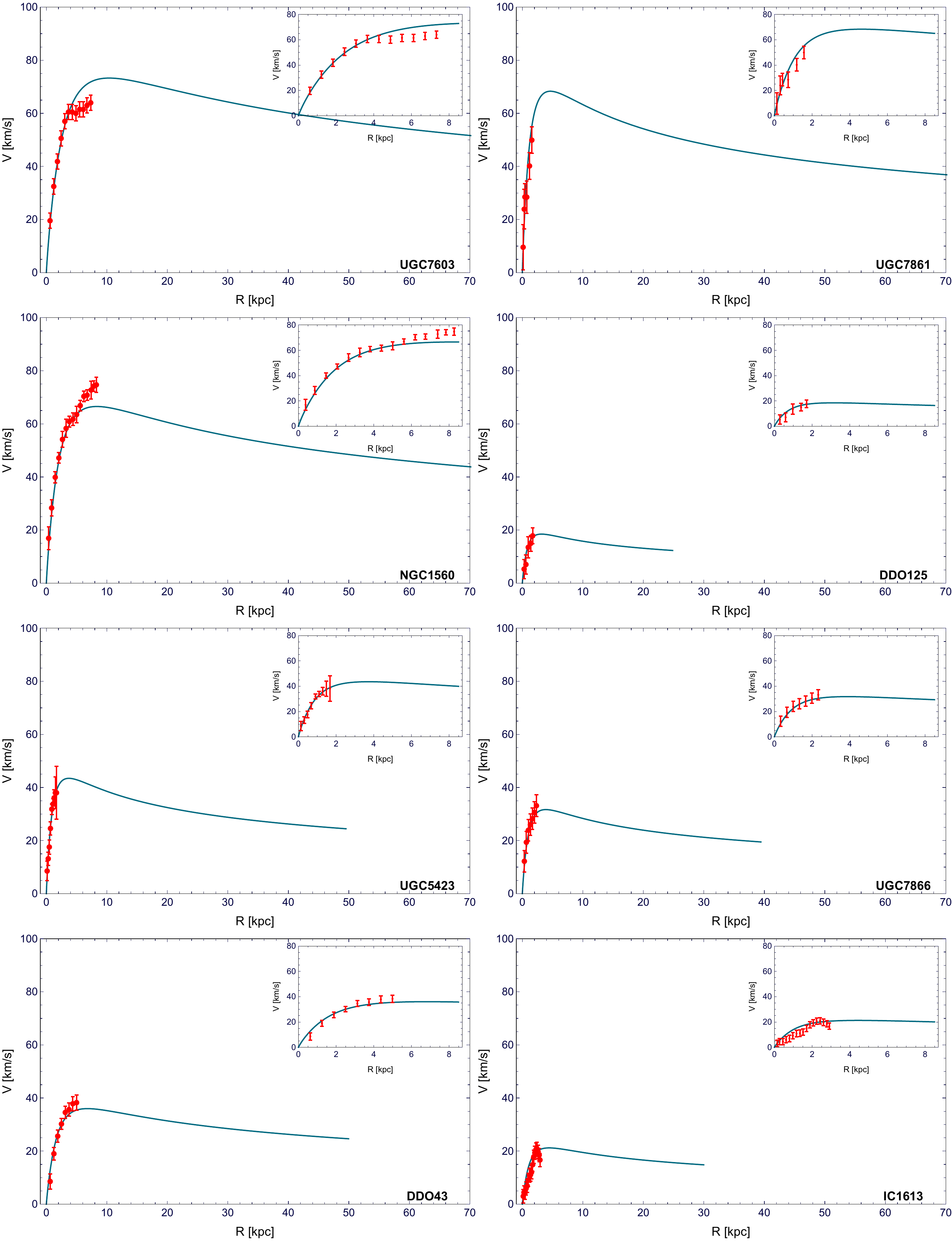}
\contcaption{\footnotesize{\it{continued}}}
\end{center}
\end{figure*}

\newpage

\begin{figure*}[h!]
\begin{center}
\includegraphics[angle=0,height=18.8truecm,width=12.8truecm]{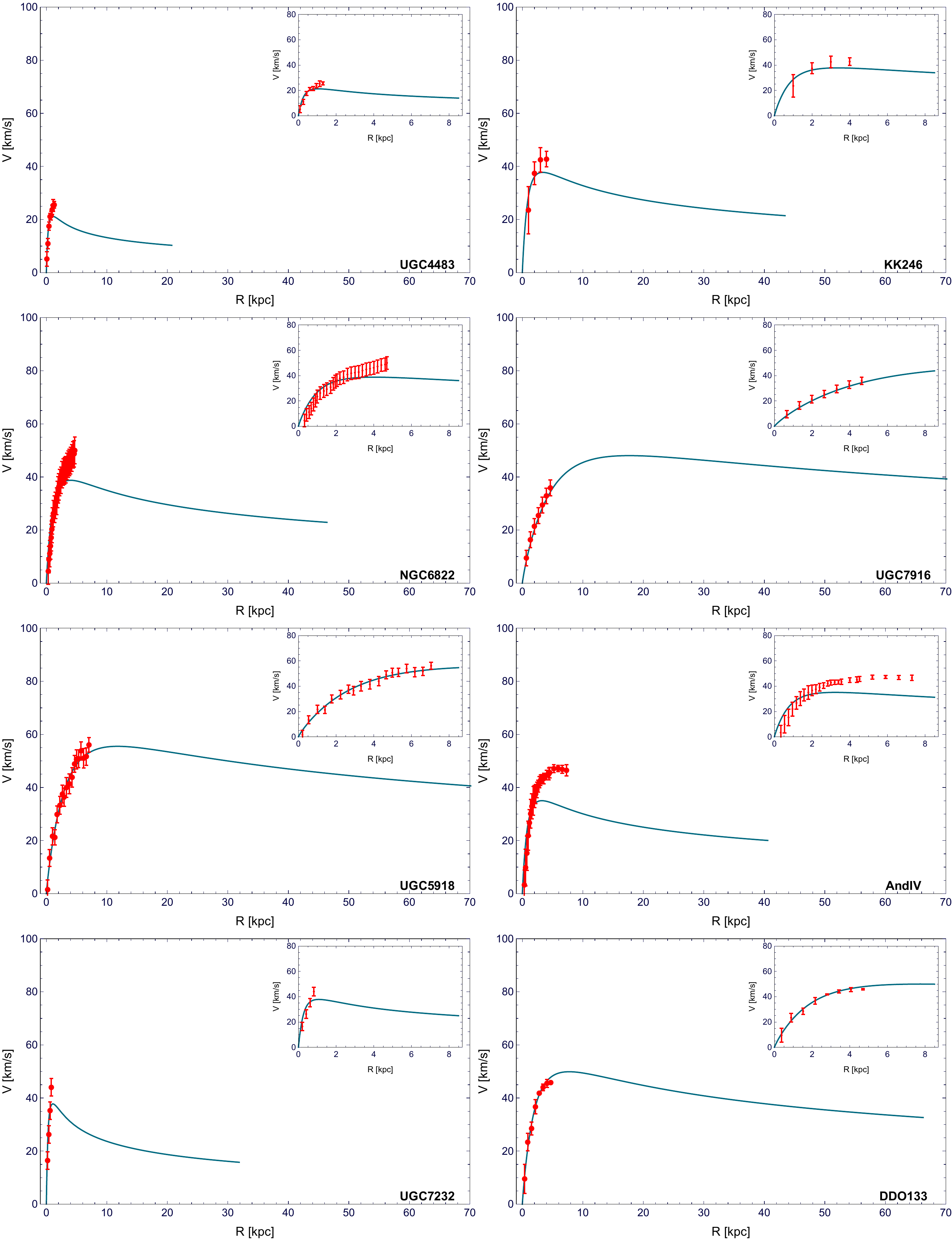}
\contcaption{\footnotesize{\it{continued}}}
\end{center}
\end{figure*}

\begin{figure*}[h!]

\begin{center}
\includegraphics[angle=0,height=9.8truecm,width=12.8truecm]{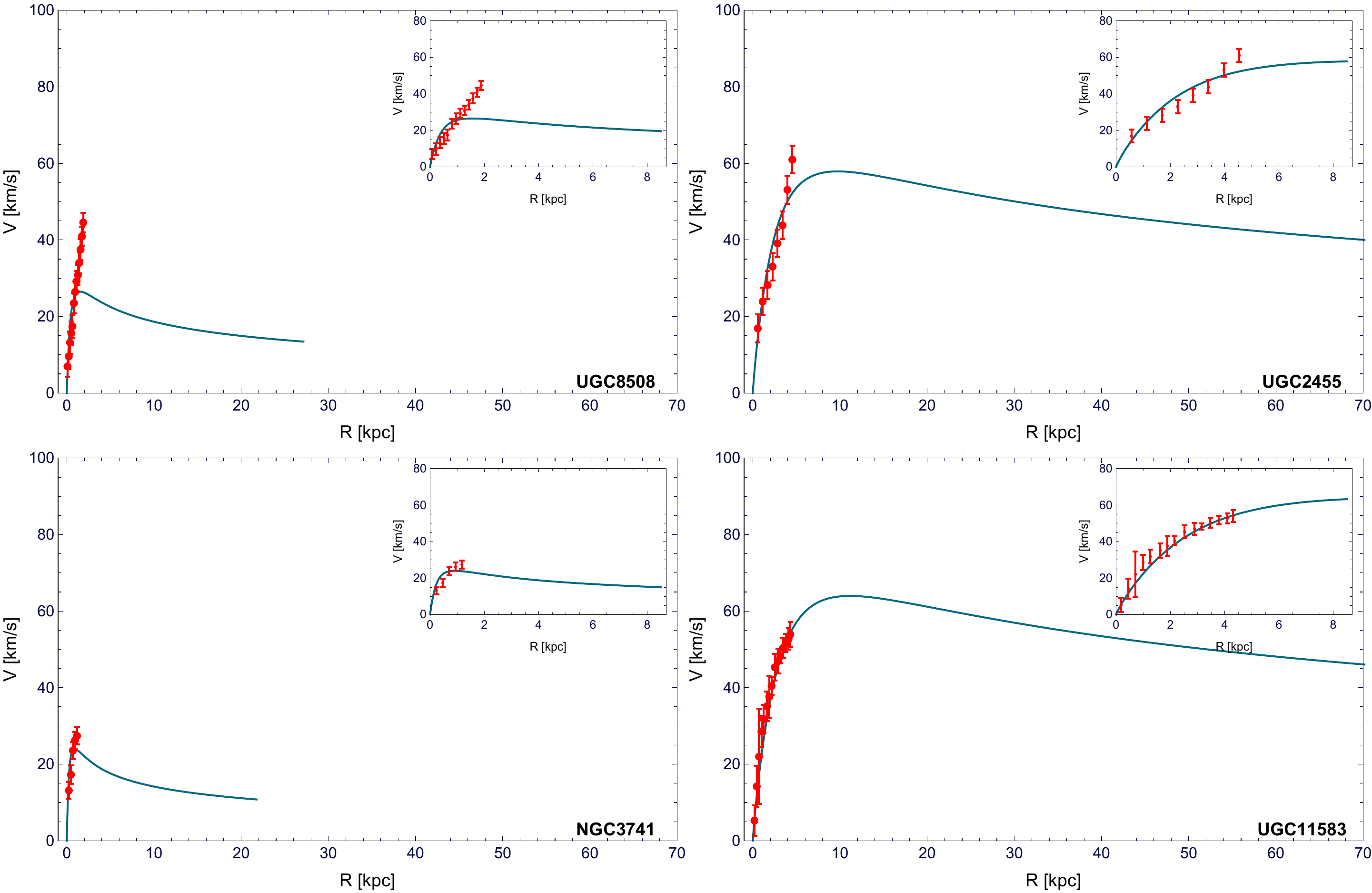}
\contcaption{\footnotesize{\it{continued}}}

\end{center}
\end{figure*}

\vspace{2.1cm}
\section{Details on the astrophysical factor}
\label{AppB}

\begin{figure*}[th!]
{\includegraphics[angle=0,height=7.5truecm,width=10.0truecm]{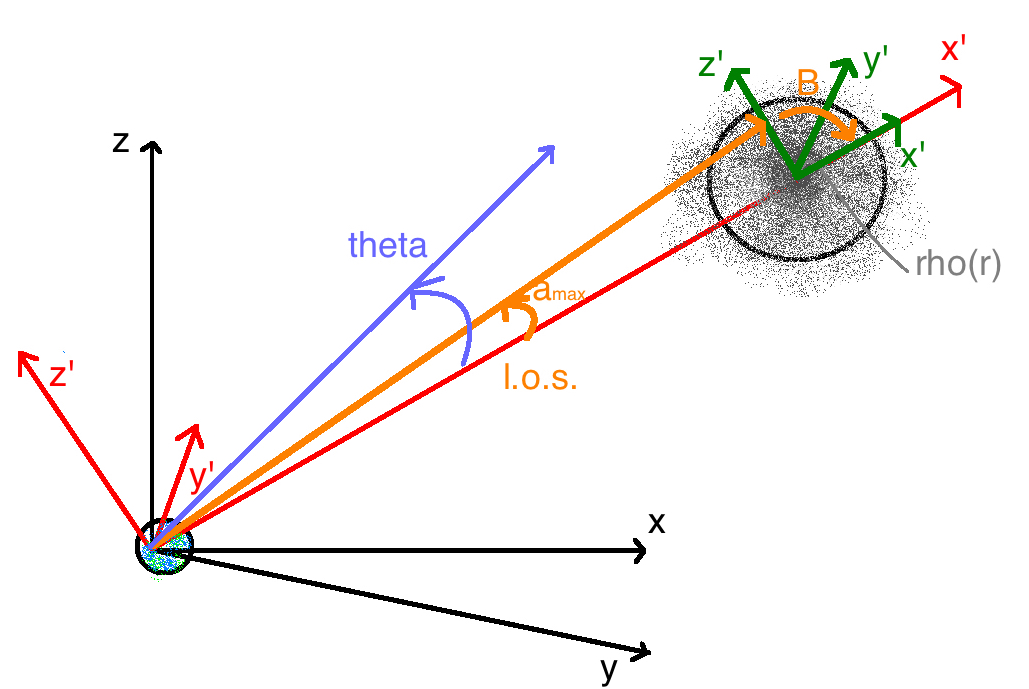}}
\caption {\centerlastline \footnotesize Coordinate reference systems for the J-factor estimation. The Earth is fixing the position of the observer, regardless of whether it is a ground-based or satellite telescope. Following Eq. (\ref{changecoord}): $\rho(r)$ (gray) is the DM density distribution profile in spherical coordinates centred on the target reference system. Green axes ($x',y',z'$) represent the cartesian reference frame centred on the source. Here, one of the axes is aligned with the $l.o.s.$. The red axes ($x'-d,y',z'$) are the Earth centred cartesian coordinate system. Finally, the orange axes ($l,\alpha,\beta$) represent the $l.o.s.$ coordinates from the observer to the source. The blue line ideally shows the angle $\theta$, that is defined by the instrument PSF for point-like analysis.  $\hspace{9cm}$}
\label{coord}

\end{figure*}

Although the astrophysical J-factor is extensively used in indirect DM searches,  
estimating its value is sometimes not straightforward, unless to spend a time to well understand it.
Typically, the DM density distribution in the selected source is given
by the spherical coordinates $\rho(r)$, in the target reference frame. Calculating the J-factor requires to move from the source to the observer reference system. Then, it is straightforward to recover the required coordinate reference system along the $l.o.s.$ as follows:

\begin{equation}
\rho(r)\equiv\rho(x',y',z')\equiv\rho(x'-d,y',z')\equiv\rho(l,\alpha,\beta).
\label{changecoord}
\end{equation}  

\noindent
Here, $r$ and $(x',y',z')$ are the radial and the cartesian coordinates, respectively, in the reference system centred on the target. On the other hand, $(x,y,z)\equiv(x'-d,y',z')$ and $(l,\alpha,\beta)$ represent the cartesian and the $l.o.s.$ coordinates in the observer reference frame. Changing the reference system as:

\begin{eqnarray}
(x',y',z')_\mathrm{target}\equiv(x'-d,y',z')_\odot \equiv  \nonumber \\
 \quad \equiv(l\cos\alpha -d,l\sin\alpha\sin\beta,l\sin\alpha\cos\beta)
\end{eqnarray}

\noindent
the radius as a function of the l.o.s. coordinates is given by:

\begin{equation}
r^2=x'^2+y'^2+z'^2=d^2-2dl\cos\alpha+l^2.
\label{los}
\end{equation}

\noindent
Then, the integration limits are:

\begin{equation}
l_\mathrm{min/max}(r,d,\alpha)=d\cos\alpha\pm\sqrt{r^2-d^2\sin^2\alpha}\,\, .
\end{equation}

\noindent
If $\alpha=0$, then $l_\mathrm{min/max}=d\pm r$. Here, $r=r_\mathrm{max}$ is the virial (rotationally-supported objects) or tidal (pressure-supported objects) radius, depending on the kind of galaxy that has been previously selected as a target. If $\alpha\neq0$, then $r^2\gsim d^2\sin^2\alpha$. This also gives the angular versus the radial dimensions of the source in the sky as $\alpha=\arcsin(r/d)$. As discussed in the main text, if $r_\mathrm{max}=R_\mathrm{vir}$, the projected virial angular dimension of the source in sky is $\alpha_\mathrm{vir}$. Geometrically speaking, if the angular resolution of the instrument is $\theta_\mathrm{PSF}\lsim\alpha_\mathrm{target}$ the source can be resolved by the instrument. Otherwise, if $\theta_\mathrm{PSF}\gsim\alpha_\mathrm{target}$, the source is unresolved.

\section{Individual galaxy analysis }
\label{AppC}

Upper limits on the DM particle mass and annihilation cross-section from the analysis of the individual 36 dIrr galaxies.

\begin{figure*}[h!]
\centering
\includegraphics[angle=0,height=10truecm,width=18truecm]{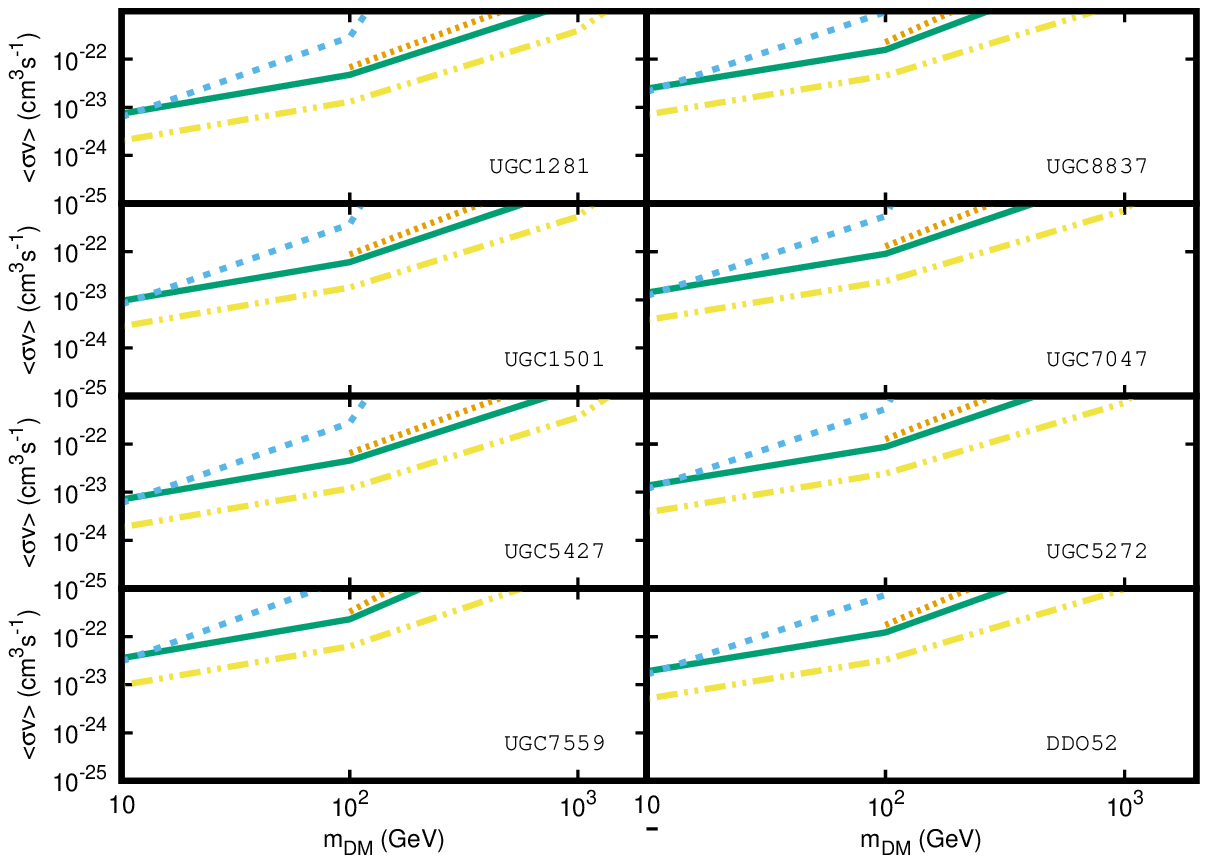}

\caption{\centerlastline \footnotesize{Upper limits on the DM particle mass and annihilation cross-section given by the individual galaxy analysis with statistical significance of $3\sigma$, angular resolution of $0.5^\circ$ and three annihilation channels: $b\bar b$ (full-green line), $\tau^+\tau^-$ (dashed-blue line) and $W^+W^-$ (dotted-orange line). The dashed-dotted-yellow line shows the constraints with a PSF of $0.1^\circ$ for the $b\bar b$ annihilation channel.  $\hspace{9.2cm}$}}
\label{1Multi}
\end{figure*}

\begin{figure*}
\centering
\includegraphics[angle=0,height=10truecm,width=18truecm]{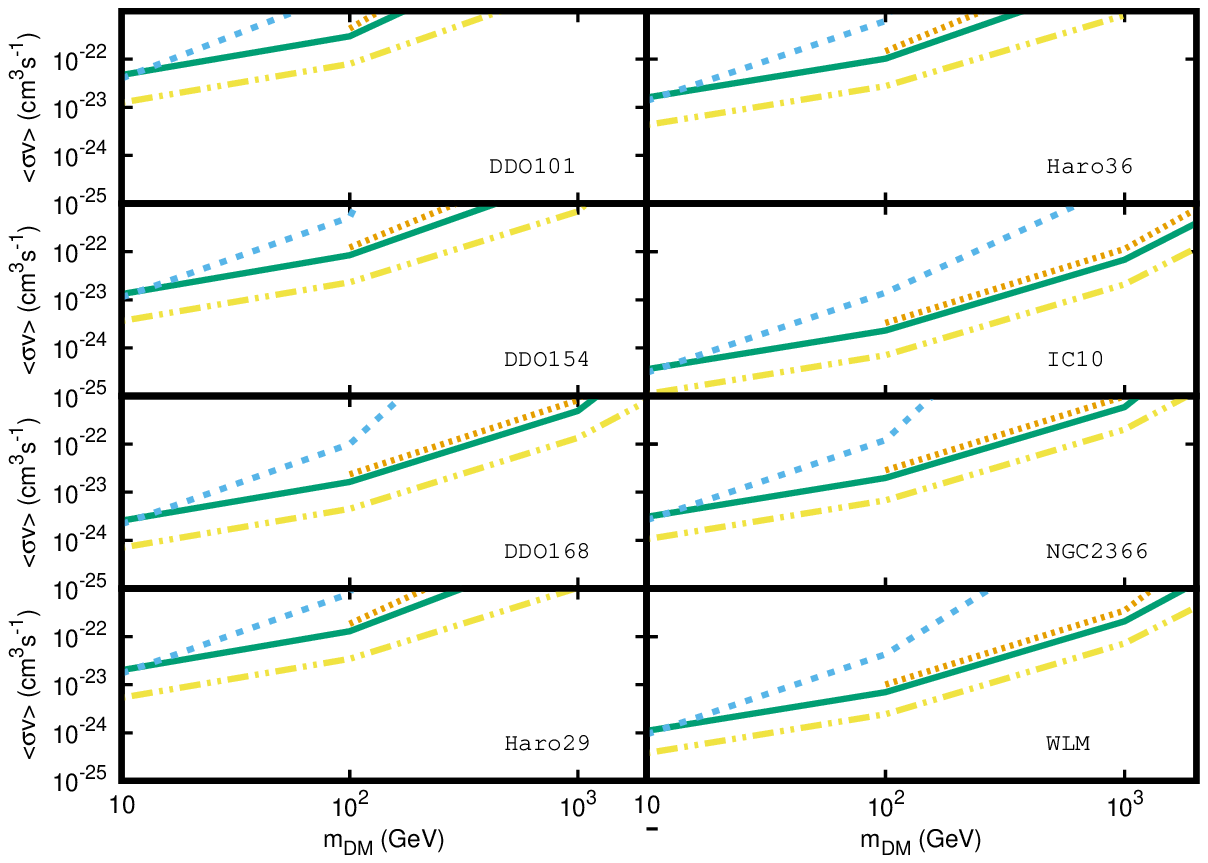}
\includegraphics[angle=0,height=10truecm,width=18truecm]{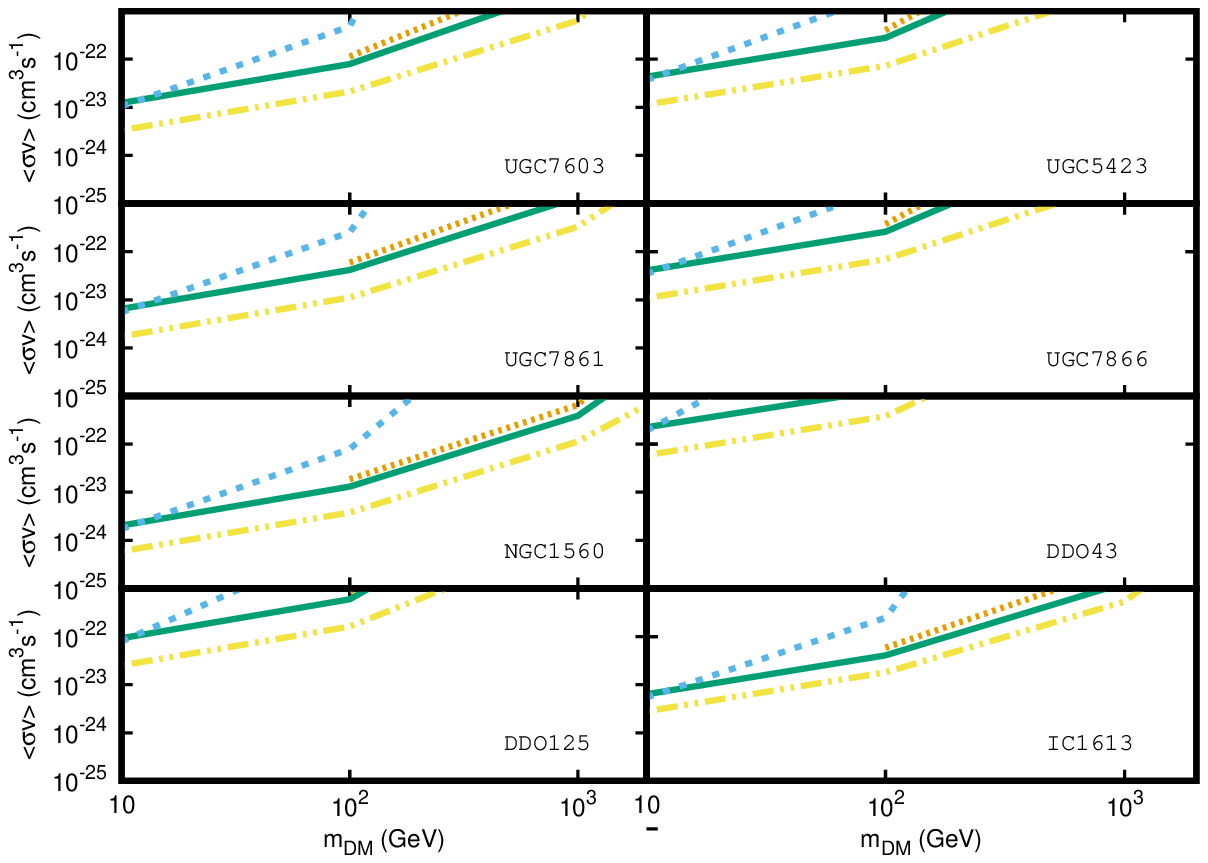}

\contcaption{\footnotesize{\it{continued}}}
\label{23Multi}
\end{figure*}

%\onecolumn
\begin{figure*}

\centering
\includegraphics[angle=0,height=10truecm,width=18truecm]{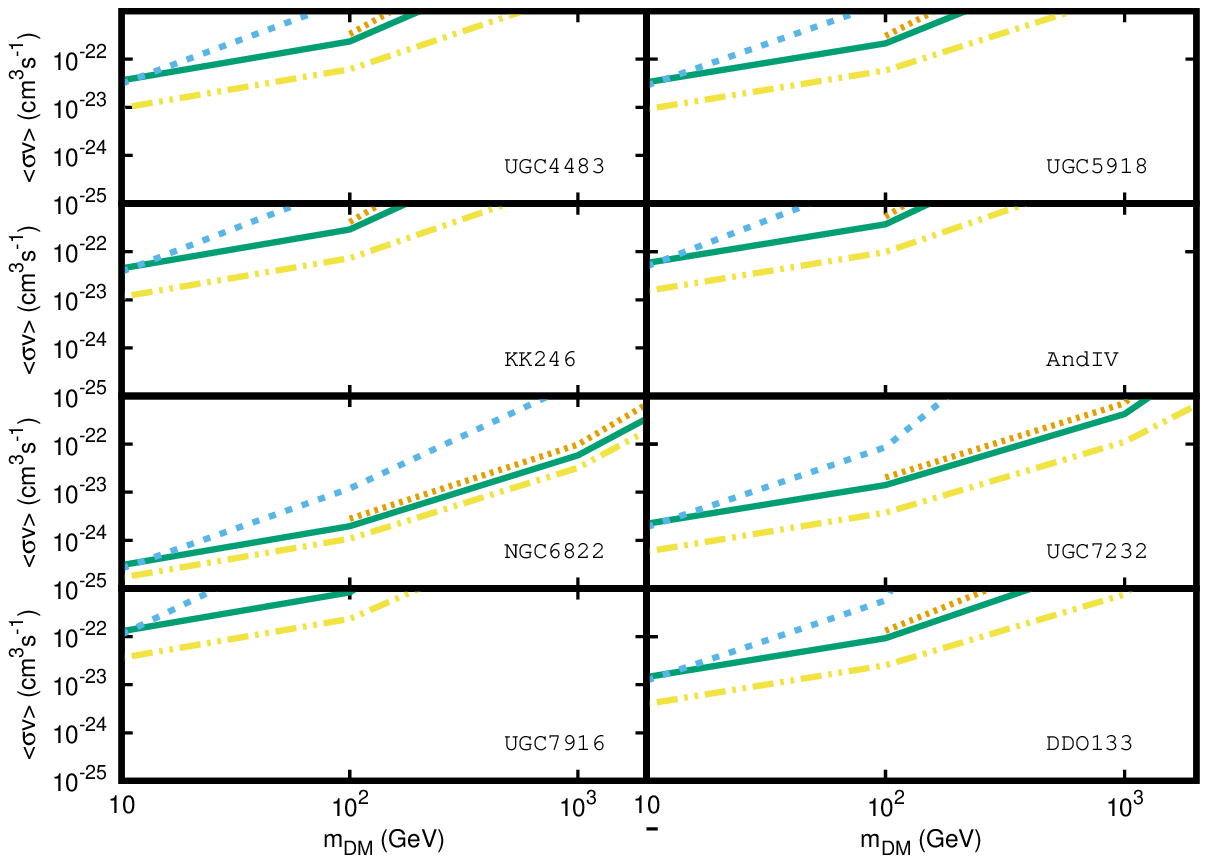}
\includegraphics[angle=0,height=10truecm,width=18truecm]{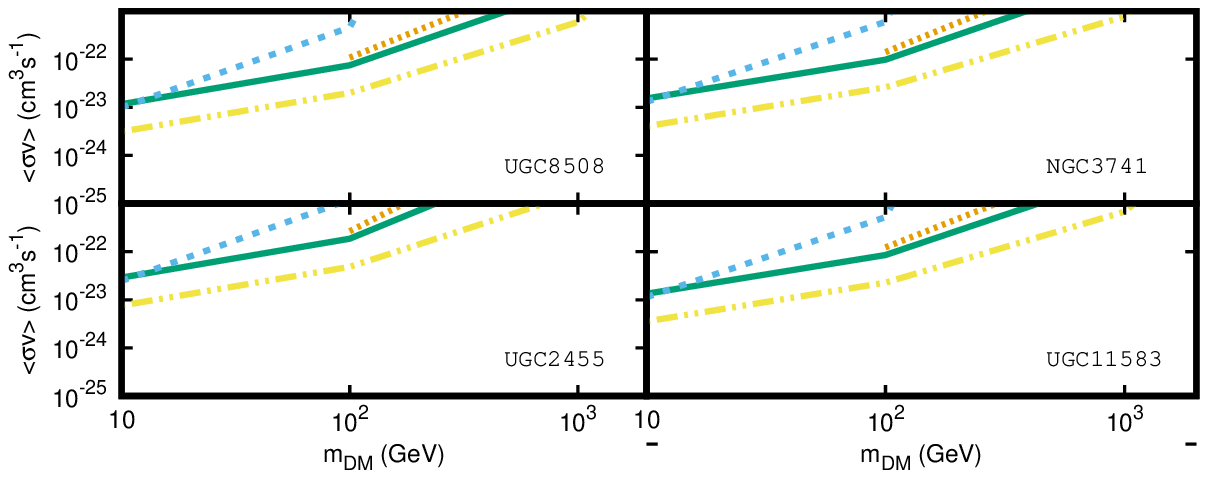}

\contcaption{\footnotesize{\it{continued}}}
\label{45Multi}
\end{figure*}

%%%%%%%%%%%%%%%%%%%%%%%%%%%%%%%%%%%%%

%\newpage

%\nocite{*}

%\bibliography{biblio}% Produces the bibliography via BibTeX.

\end{document}